\documentclass[aps,prd,onecolumn,superscriptaddress,nofootinbib,longbibliography]{revtex4-2}

\usepackage[utf8]{inputenc}
\usepackage[T1]{fontenc}
\usepackage{amsmath,amssymb,amsfonts,bm}
\usepackage{graphicx}
\usepackage[colorlinks=true,linkcolor=blue,citecolor=blue,urlcolor=blue]{hyperref}
\usepackage{float}
\usepackage{booktabs}
\usepackage{multirow}
\usepackage{array}

\begin{document}

\title{Propagation of Laguerre-Gaussian and Bessel-Gaussian scalar beams in an effective anisotropic background}

\author{C. A. Escobar}
\email[Corresponding author: ]{carlos.escobar@xanum.uam.mx}
\affiliation{Departamento de F\'isica, Universidad Aut\'onoma Metropolitana-Iztapalapa, San Rafael Atlixco 186, 09340 Ciudad de M\'exico, M\'exico}

\author{Rom\'an Linares}
\email{lirr@xanum.uam.mx}
\affiliation{Departamento de F\'isica, Universidad Aut\'onoma Metropolitana-Iztapalapa, San Rafael Atlixco 186, 09340 Ciudad de M\'exico, M\'exico}

\author{E. Pl\'acido-Flores}
\email{lalo@xanum.uam.mx}
\affiliation{Departamento de F\'isica, Universidad Aut\'onoma Metropolitana-Iztapalapa, San Rafael Atlixco 186, 09340 Ciudad de M\'exico, M\'exico}

\begin{abstract}
We investigate the propagation of structured scalar optical beams in an effective anisotropic background inspired by the scalar sector of the Standard-Model Extension and controlled by a single dimensionless parameter $\lambda$. The physically relevant configuration is a transverse radial director field that modifies the radial part of the Helmholtz operator while preserving axial symmetry. Starting from the Green-function representation, we cast the propagation problem as an initial-value spectral reconstruction of a prescribed finite-aperture entrance profile at $z=0$ and verify that this profile is recovered at the launch plane across the values of $\lambda$ used in the analysis, within small numerical error. We use the Laguerre--Gaussian mode $L_3$ as the representative vortex-free Laguerre case, retain $L_4$ only as a quantitative benchmark for radial-order dependence, and compare both with a Bessel--Gaussian beam of input order $m=0$. The effective anisotropy produces a systematic redistribution of radial intensity, determines whether the central peak remains dominant or is overtaken by off-axis maxima as propagation advances, and controls the radial displacement of the dominant side lobes. For the finite-aperture Bessel--Gaussian beam, the same parameter quantifies how the approximately diffraction-resistant ring structure broadens for negative $\lambda$ and compresses for positive $\lambda$ during propagation.
\end{abstract}
\maketitle

\section{Introduction}

Structured light provides a versatile platform for controlling the spatial, phase, and polarization content of optical fields. Laguerre--Gaussian (LG) and Bessel--Gaussian (BG) beams are among the most widely used families because they combine experimentally accessible field profiles with well-defined radial structure, orbital-angular-momentum content, and robust propagation signatures \cite{allen1992orbital,mair2001entanglement,wang2012terabit,Mazilu2010,Forbes2021Structured,padgett2017orbital}. These properties have stimulated applications in free-space communication, optical manipulation, metrology, microscopy, and topological light engineering.

Structured beams have also become useful probes of effective geometries, anisotropic media, and synthetic gauge effects \cite{Longhi2010Dirac,Bliokh2015SpinOrbit,Thompson2010Dielectric,Bekenstein2017Control}. In genuinely anisotropic optical settings, however, propagation depends on the orientation of the optical axis, the tensorial dielectric response, and, in general, on polarization. Vector Bessel beams in bianisotropic media, cylindrically symmetric propagation in uniaxial crystals, LG/BG beams in uniaxial crystals, and vector-vortex Bessel--Gauss beams are representative examples of this broader literature \cite{Novitsky2005BesselBianisotropic,Ciattoni2002Cylindrical,Cincotti2002Uniaxial,Huang2011VectorVortexBG,Egorov2024Uniaxial,Berskys2022SphericalVBessel}. Recent work on polarization singularities and topological textures has further highlighted how structured beams respond to anisotropy and free-space propagation in experimentally accessible situations \cite{Zhen2025Cpoint,Zhen2026Skyrmion}.

The present manuscript addresses a different, deliberately reduced problem. We work with a scalar effective propagation model inspired by the scalar sector of the Standard-Model Extension (SME), not with a full vectorial Maxwell treatment of a crystal or a bianisotropic slab. The SME is an effective-field-theory framework that parametrizes controlled departures from exact Lorentz invariance through background coefficients coupled to otherwise conventional fields, thereby providing a systematic language for studying anisotropic modifications of wave propagation~\cite{PhysRevD.55.6760,PhysRevD.58.116002,Kostelecky:2008ts}. Accordingly, the parameter $\lambda$ introduced below should not be interpreted as a replacement for a dielectric tensor; rather, it plays the role of a reduced radial-anisotropy parameter after projection onto an axially symmetric scalar sector. The comparison with optical anisotropy is therefore qualitative: the purpose is not to fit a specific crystal platform, but to identify how an effective radial anisotropy reorganizes structured-beam propagation once a finite profile is prescribed at a launch plane. From this perspective, the novelty of the present analysis is not a crystal-specific constitutive model but a nonparaxial scalar-effective treatment that keeps the launch-plane profile fixed and quantifies how LG and BG beams redistribute weight among competing radial channels as $\lambda$ varies. This scalar-effective treatment can be regarded as an approximation to a real optical beam whenever the relevant propagation dynamics is dominated by a single transverse amplitude, so that polarization mixing, birefringence, and other genuinely vectorial effects may be neglected at leading order.

Our starting point is the massless scalar sector of the SME ~\cite{PhysRevD.55.6760,PhysRevD.58.116002}, considered here in the scalar-field context discussed in Ref.~\cite{EdwardsKostelecky2018FinslerScalar} and in subsequent Casimir and effective-geometry applications \cite{Cruz2017CasimirLV,Escobar2020CasimirLocal,Escobar2020NonPerturbative,MartinRuiz2020Sphere,EscobarRuiz2021Cylinder,BezerraDeMello2023ScalarLV}. In unbounded space, some Lorentz-violating coefficients can be removed by field or coordinate redefinitions \cite{Kostelecky:2010ze,Borges:2018fso}; however, once a beam profile is prescribed at a launch plane, the propagation problem becomes nontrivial and acquires observable dependence on the anisotropy parameter.

The main goals of this work are threefold. First, we reformulate the propagation problem in terms of an initial-value
spectral reconstruction adapted to the same axisymmetric propagation problem
described by the Green-kernel formulation, and verify numerically at
$\tilde{z}=0$ that the prescribed entrance profile is recovered across the
set of $\lambda$ values used in the numerical study. Second, we quantify the anisotropy-induced redistribution of radial intensity through observables such as the dominant-peak position, central intensity, effective radius, selected width measures, and the axial onset of off-axis dominance. Third, we compare LG and BG inputs within the same scalar-effective framework and identify which features are generic and which depend on the beam family.

The paper is organized as follows. Section \ref{Model1} introduces the effective scalar model, clarifies its relation to optical anisotropy, and presents the spectral initial-value implementation used for the numerical results. Section \ref{Prop1} discusses LG and BG propagation, with $L_3$ as the representative LG mode, $L_4$ as a quantitative benchmark for radial-order dependence, and BG as the complementary finite-energy Bessel case. Section \ref{Conclu1} summarizes the physical picture that emerges from the effective-anisotropy parameter $\lambda$. Appendix~\ref{App1} clarifies the relation between the working spectral reconstruction
and the axisymmetric Green-kernel formulation once the problem is posed in
terms of a prescribed launch-plane profile.

\section{Effective scalar model and spectral implementation}
\label{Model1}

\subsection{Lorentz-violating scalar operator}

Our starting point is the massless scalar sector of the minimal SME, described by the Lagrangian density
\begin{equation}
\mathcal{L}_{\phi}=\frac{1}{2}(\partial^{\mu}\phi)(\partial_{\mu}\phi)+\frac{1}{2}h^{\mu\nu}(\partial_{\mu}\phi)(\partial_{\nu}\phi),
\label{eq:lagrangian}
\end{equation}
where $h^{\mu\nu}$ is a constant symmetric background tensor. We consider the parametrization
\begin{equation}
h^{\mu\nu}=\lambda u^{\mu}u^{\nu},
\label{eq:hmunu}
\end{equation}
with $u^{\mu}=(u_0,\mathbf{u})$, $u_\mu u^\mu=1$, and dimensionless anisotropy strength $\lambda$. The Euler--Lagrange equation becomes
\begin{equation}
\left[\Box+\lambda (u\!\cdot\!\partial)^2\right]\phi(x)=0.
\label{eq:eom}
\end{equation}
We restrict $-1<\lambda<1$ so that the effective operator remains hyperbolic and the radial kinetic term does not change sign. This interval is also consistent with standard SME phenomenological bounds on Lorentz-violating coefficients \cite{Kostelecky:2008ts}.

For monochromatic fields $\phi(x)=\phi_{\omega}(\mathbf{r})e^{-i\omega t}$, the Green function satisfies
\begin{equation}
\left\{(\omega/c)^2+\nabla^2+\lambda\big[-iu_0(\omega/c)-\mathbf{u}\!\cdot\!\nabla\big]^2\right\}G_u(\mathbf{x},\mathbf{x}')=-\delta^{(3)}(\mathbf{x}-\mathbf{x}').
\label{eq:greenDef}
\end{equation}
The timelike choice $u^{\mu}=(1,\mathbf{0})$ only rescales the frequency and does not modify the transverse profile, so we focus on the physically relevant radial spacelike configuration.

\subsection{Radial spacelike background as an effective anisotropy}

We take the preferred spatial direction to be locally radial in the transverse plane. In cylindrical coordinates $(\rho,\theta,z)$ this gives
\begin{equation}
\left[(\omega/c)^2+\frac{\partial^2}{\partial z^2}+(1-\lambda)\frac{\partial^2}{\partial \rho^2}+\frac{1}{\rho}\frac{\partial}{\partial \rho}+\frac{1}{\rho^2}\frac{\partial^2}{\partial\theta^2}\right]G_u(\rho,z;\rho',z')=-\frac{1}{\rho}\delta(\rho-\rho')\delta(z-z')\delta(\theta-\theta').
\label{eq:radialGreenEq}
\end{equation}
The corresponding Green function can be written as \cite{Escobar2025}
\begin{equation}
G_u(\rho,z;\rho',z')=\frac{i\pi}{1-\lambda}\left(\frac{\rho}{\rho'}\right)^{\frac{\lambda}{2(\lambda-1)}}
\sum_{m=-\infty}^{\infty}\int_{-\infty}^{\infty}\frac{dk}{(2\pi)^2}e^{ik(z-z')}e^{im(\theta-\theta')}J_{n_m}(\chi\rho_<)H^{(1)}_{n_m}(\chi\rho_>),
\label{eq:greenKernel}
\end{equation}
where $\rho_< = \min(\rho,\rho')$, $\rho_> = \max(\rho,\rho')$,
\begin{equation}
\chi=\frac{\sqrt{(\omega/c)^2-k^2}}{\sqrt{1-\lambda}},\qquad
n_m=\frac{\sqrt{4m^2(1-\lambda)+\lambda^2}}{2(1-\lambda)}.
\label{eq:chi_nm}
\end{equation}
In the axially symmetric sector used below, $m=0$ and the effective order is
\begin{equation}
\nu_0\equiv n_0=\frac{|\lambda|}{2(1-\lambda)}.
\label{eq:nu0}
\end{equation}

Equation~\eqref{eq:radialGreenEq} should therefore be read as the axisymmetric scalar reduction of an anisotropic propagation problem. In this setting, the transverse radial director field selects the radial channel, and the parameter $\lambda$ modifies the relative weight of radial derivatives in the effective Helmholtz operator. The resulting dynamics preserves axial symmetry and produces controlled radial reshaping, but leaves polarization-dependent effects outside the present scope. The model is therefore best viewed as complementary to fully vectorial crystal-specific descriptions.

\subsection{Initial-value spectral representation}

The Green-function kernel in Eq.~\eqref{eq:greenKernel} provides the formal solution of the
axisymmetric propagation problem. For the numerical analysis, however, it is
more convenient to work with an initial-value representation adapted to a
prescribed entrance profile at $\tilde z = 0$. In this representation the
field is expanded over regular radial modes, while the corresponding spectral
amplitude is fixed by the launch-plane data. Thus the spectral reconstruction
used below addresses the same axisymmetric propagation problem as the
Green-kernel formulation, but in the form most suitable for imposing
prescribed launch-plane data. For completeness, the relation between both
descriptions is summarized in Appendix~\ref{App1}.

Using the dimensionless variables
\begin{equation}
\tilde\rho=\rho/\rho_0,\qquad \tilde z=z/\rho_0,\qquad \Omega=\omega\rho_0/c,\qquad \kappa=\rho_0/w_0,
\label{eq:dimensionless}
\end{equation}
we define
\begin{equation}
\sigma=\frac{\lambda}{2(\lambda-1)},
\qquad
\tilde\beta(\eta)=\sqrt{\Omega^2-(1-\lambda)\eta^2},
\label{eq:sigma_beta}
\end{equation}
and the branch-adapted radial basis
\begin{equation}
\mathcal{J}_{\lambda}(x)=
\begin{cases}
J_{\nu_0}(x), & \lambda\ge 0,\\[4pt]
J_{-\nu_0}(x), & \lambda< 0.
\end{cases}
\label{eq:Jlambda}
\end{equation}
For $\lambda<0$ one has $\sigma=\nu_0>0$ and the small-argument limits $J_{\pm\nu_0}(x)\sim x^{\pm\nu_0}$. Therefore $\tilde\rho^{\sigma}J_{\nu_0}(\eta\tilde\rho)\sim \tilde\rho^{2\nu_0}$ vanishes on axis, whereas $\tilde\rho^{\sigma}J_{-\nu_0}(\eta\tilde\rho)\sim \tilde\rho^0$ remains finite. The branch choice in Eq.~\eqref{eq:Jlambda} is thus the one compatible with bright on-axis entrance profiles in the reduced scalar problem.

For an axially symmetric input amplitude $\Phi_{\mathrm{in}}(\tilde\rho)$ prescribed at $\tilde z=0$, the propagated field is reconstructed as
\begin{equation}
\phi(\tilde\rho,\tilde z)=\tilde\rho^{\sigma}\int_0^{\infty}d\eta\;\eta\,e^{i\tilde\beta(\eta)\tilde z}\,\mathcal{J}_{\lambda}(\eta\tilde\rho)\,\mathcal{A}(\eta),
\label{eq:spectralPhi}
\end{equation}
with spectral amplitude
\begin{equation}
\mathcal{A}(\eta)=\int_0^1 d\tilde\rho'\;\tilde\rho'^{\,\sigma+\frac{1}{1-\lambda}}\mathcal{J}_{\lambda}(\eta\tilde\rho')\,\Phi_{\mathrm{in}}(\tilde\rho').
\label{eq:spectralA}
\end{equation}
The associated intensity is $I(\tilde\rho,\tilde z)=|\phi(\tilde\rho,\tilde z)|^2$.

Equations~\eqref{eq:spectralPhi}--\eqref{eq:spectralA} are the working formulas for all datasets used in this manuscript. Once the problem is posed in terms of a prescribed entrance profile, the spectral representation provides the most direct way to propagate that common initial state across different values of $\lambda$.

\subsection{Numerical implementation and entrance-plane validation}

The numerical integrations are performed with nonuniform meshes in $\eta$, $\tilde\rho$, and $\tilde z$. The $\eta$ grid is concentrated around the transition scale $\eta_c=\Omega/\sqrt{1-\lambda}$, where $\tilde\beta(\eta)$ varies most rapidly. The radial input mesh is refined near the aperture edge $\tilde\rho=1$, and the propagation grid is sampled more densely close to the launch plane in order to resolve the early stages of the evolution. In all results we set
\begin{equation}
\Omega=1,\qquad \kappa=1,
\label{eq:omegaKappaChoice}
\end{equation}
while for BG beams we use $\tilde k_r=k_r\rho_0=5.9$, which yields a clearly resolved central lobe and first ring within the normalized radial window.

The finite aperture is implemented through the common support $0\le \tilde\rho\le 1$. This is important for interpreting the entrance-plane structure of the truncated Laguerre inputs: within the aperture, $L_3(2\tilde\rho^2)$ has one radial zero, whereas $L_4(2\tilde\rho^2)$ has two. The truncated $L_3$ and $L_4$ entrance profiles are therefore already distinct at $\tilde z=0$, without implying any inconsistency in the propagation scheme. For each beam family, we normalize the intensity by the peak value of the prescribed entrance profile,
\begin{equation}
I_0^{(\mathrm{fam})}\equiv \max_{\tilde\rho}\,|\Phi_{\mathrm{in}}^{(\mathrm{fam})}(\tilde\rho)|^2,
\end{equation}
so that all panels corresponding to the same family share a common reference intensity independent of $\lambda$.

To quantify the entrance-plane reconstruction we compare the reconstructed intensity
\begin{equation}
I_{\rm rec}(\tilde\rho)=|\phi(\tilde\rho,0)|^2
\end{equation}
with the prescribed reference intensity
\begin{equation}
I_{\rm ref}(\tilde\rho)=|\Phi_{\mathrm{in}}(\tilde\rho)|^2.
\end{equation}
The relative $L^2$ error reported below is the root-mean-square discrepancy between $I_{\rm rec}$ and $I_{\rm ref}$, normalized by the $L^2$ norm of $I_{\rm ref}$,
\begin{equation}
\epsilon_{L^2}
=
\frac{\left(\int_0^1 d\tilde\rho\,\big|I_{\rm rec}(\tilde\rho)-I_{\rm ref}(\tilde\rho)\big|^2\right)^{1/2}}
{\left(\int_0^1 d\tilde\rho\,|I_{\rm ref}(\tilde\rho)|^2\right)^{1/2}}.
\label{eq:L2error}
\end{equation}
We also monitor the integrated leakage outside the aperture,
\begin{equation}
\mathcal L_{\rm out}=\int_1^{2.0} I(\tilde\rho,0)\,d\tilde\rho,
\label{eq:Lout}
\end{equation}
which measures the residual reconstructed intensity lying beyond the prescribed support in the entrance-plane validation window.

\begin{table}[t]
\centering
\caption{Validation of the spectral reconstruction at $\tilde z=0$ for the truncated Laguerre--Gaussian inputs. We list the relative $L^2$ error of the intensity profile and the integrated leakage outside the aperture, $\mathcal L_{\rm out}=\int_1^{2.0} I(\tilde\rho,0)\,d\tilde\rho$.}
\label{tab:z0validation}
\begin{tabular}{cccc}
\toprule
$p$ & $\lambda$ & intensity error (\%) & $\mathcal L_{\rm out}$ \\
\midrule
3 & -0.35 & 0.53 & $1.0\times 10^{-5}$ \\
3 & 0.00  & 0.69 & $1.0\times 10^{-5}$ \\
3 & 0.35  & 1.42 & $1.0\times 10^{-5}$ \\
3 & 0.50  & 2.32 & $1.0\times 10^{-5}$ \\
4 & -0.35 & 0.50 & $9.7\times 10^{-6}$ \\
4 & 0.00  & 0.69 & $9.6\times 10^{-6}$ \\
4 & 0.35  & 1.68 & $9.5\times 10^{-6}$ \\
4 & 0.50  & 3.26 & $9.6\times 10^{-6}$ \\
\bottomrule
\end{tabular}
\end{table}

Table~\ref{tab:z0validation} quantifies the reconstruction error with respect to the prescribed reference profile. Separately, for the full symmetric datasets, the maximum pairwise relative $L^2$ difference among distinct $\lambda$ values at $\tilde z=0$ remains below $2.1\times10^{-2}$ for $L_3$, below $3.0\times10^{-2}$ for $L_4$, and below $1.1\times10^{-2}$ for BG. The differences discussed in the propagation figures therefore arise from the subsequent anisotropy-dependent evolution and not from mismatched launch conditions.

For quantitative comparisons we report the dominant-peak position $\tilde\rho_{\rm peak}$, the corresponding intensity $I_{\rm peak}$, the on-axis value $I_{\rm center}=I(0,\tilde z)$, and the first radial minimum $\tilde\rho_{\rm min}$. Here $\tilde\rho_{\rm peak}$ denotes the radial position of the dominant intensity maximum at fixed $\tilde z$: when the on-axis peak remains dominant one has $\tilde\rho_{\rm peak}=0$, whereas once an off-axis maximum becomes dominant, $\tilde\rho_{\rm peak}$ records its radial location. We also evaluate an effective radius defined from the second radial moment over an extended observation window,
\begin{equation}
 r_{\rm eff}^2(\tilde z)=\frac{\int_0^{4.0} d\tilde\rho\;\tilde\rho^3 I(\tilde\rho,\tilde z)}{\int_0^{4.0} d\tilde\rho\;\tilde\rho\, I(\tilde\rho,\tilde z)}.
\label{eq:reff}
\end{equation}
Physically, $r_{\rm eff}$ measures the overall radial extent of the intensity distribution: it is sensitive to global broadening or compression of the beam even when the dominant maximum remains on axis. The wider cutoff is used because the beam profiles broaden during propagation, and a smaller window underestimates the second moment at larger $\tilde z$. We verified the stability of this observable by increasing the cutoff from $\tilde\rho_{\max}=4.0$ to $4.5$: the resulting change in $r_{\rm eff}$ remains below $1.5\%$ at the representative intermediate plane $\tilde z=0.441$ for all beam families and $\lambda$ values considered, and stays below $5\%$ even at $\tilde z=1.2$. We therefore adopt $\tilde\rho_{\max}=4.0$ as a numerically stable observation window for the second-moment analysis.

For BG beams we also use the central-lobe full width at half maximum,
\begin{equation}
\mathrm{FWHM}(\tilde z)=2\tilde\rho_{1/2},\qquad I(\tilde\rho_{1/2},\tilde z)=\tfrac12 I(0,\tilde z),
\label{eq:fwhm}
\end{equation}
where $\tilde\rho_{1/2}$ is the smallest positive solution before the first radial minimum. Even when the global maximum later moves off axis, this definition continues to characterize the width of the central lobe itself.

To monitor the onset of off-axis dominance, we also define a crossover distance $\tilde z_\times$ as the first propagation point for which $\tilde\rho_{\rm peak}>0.05$; this threshold is well above the residual near-axis jitter of the reconstructed profiles and well below the radii of the dominant side maxima. Physically, $\tilde z_\times$ identifies the first axial location at which the beam is no longer dominated by the on-axis peak but by an off-axis maximum. The role of $r_{\rm eff}$ and $\tilde z_\times$ is therefore complementary: the former tracks the global radial extent of the profile, whereas the latter marks the qualitative transition to off-axis dominance. When the central peak ceases to dominate, $\tilde\rho_{\rm peak}$ identifies the strongest off-axis maximum. Axial summaries extracted from the full datasets are collected in Table~\ref{tab:axialsummary}. For all $\lambda$ and for the three beam families studied, the largest on-axis intensity occurs at launch, $\tilde z=0$, while the minimum effective radius defined in Eq.~\eqref{eq:reff} is reached very close to the launch plane. The main propagation effect is therefore not an axial displacement of a best-focus plane but a $\lambda$-dependent redistribution among radial channels. In particular, the values of $\tilde z_\times$ show that the crossover to off-axis dominance occurs earlier as $\lambda$ increases, whereas BG beams delay that crossover relative to the Laguerre family and, for $\lambda=-0.35$, do not exhibit it within the computed interval.

\begin{table}[t]
\centering
\caption{Axial localization and off-axis crossover summary extracted from the full datasets. The quantity $\tilde z_{\rm min}^{(r_{\rm eff})}$ denotes the propagation distance at which the effective radius in Eq.~\eqref{eq:reff}, evaluated over the window $0\le \tilde\rho \le 4.0$, reaches its minimum, and $\tilde z_\times$ is the first distance at which the dominant maximum becomes off axis according to the criterion $\tilde\rho_{\rm peak}>0.05$. For BG with $\lambda=-0.35$, no off-axis takeover occurs within the computed interval $0\leq \tilde z \leq 1.2$.}
\label{tab:axialsummary}
\begin{tabular}{ccccc}
\toprule
family & $\lambda$ & $\tilde z_{\rm min}^{(r_{\rm eff})}$ & $\tilde z_\times$ & $\tilde\rho_{\rm peak}(\tilde z_\times)$ \\
\midrule
$L_3$ & -0.35 & 0.0000 & 0.4917 & 0.848 \\
$L_3$ & 0.00  & 0.0000 & 0.4083 & 0.775 \\
$L_3$ & 0.35  & 0.0000 & 0.3417 & 0.725 \\
$L_3$ & 0.50  & 0.0000 & 0.3083 & 0.705 \\
$L_4$ & -0.35 & 0.0075 & 0.4583 & 0.755 \\
$L_4$ & 0.00  & 0.0025 & 0.3750 & 0.680 \\
$L_4$ & 0.35  & 0.0000 & 0.3083 & 0.635 \\
$L_4$ & 0.50  & 0.0000 & 0.2833 & 0.620 \\
BG & -0.35 & 0.0175 & $>1.2$ & --- \\
BG & 0.00  & 0.0000 & 0.6750 & 0.806 \\
BG & 0.35  & 0.0000 & 0.4750 & 0.655 \\
BG & 0.50  & 0.0000 & 0.4167 & 0.615 \\
\bottomrule
\end{tabular}
\end{table}
\section{Structured-beam propagation in the effective anisotropic background}
\label{Prop1}

\subsection{Laguerre--Gaussian beams: $L_3$ as the representative case}

Laguerre--Gaussian beams provide a natural test bed for radial reshaping because their transverse structure is organized by a discrete radial order even in the absence of orbital angular momentum. In the present axially symmetric sector, the relevant information is carried entirely by the sequence of radial maxima and minima, so these modes offer a clean way to track how the effective anisotropy redistributes intensity between the central lobe and off-axis rings as propagation proceeds.

For the Laguerre--Gaussian family we restrict attention to the axially symmetric sector $m=0$ and use the input profile
\begin{equation}
\Phi_{\mathrm{in}}^{\mathrm{LG},p}(\tilde\rho)=L_p(2\kappa^2\tilde\rho^2)e^{-\kappa^2\tilde\rho^2}\,\Theta(1-\tilde\rho).
\label{eq:LGinput}
\end{equation}
The factor $\Theta(1-\tilde\rho)$ is the Heaviside step function. It enforces the finite-aperture condition at the entrance plane by truncating the input beam outside the normalized radial interval $0\le \tilde\rho\le 1$.

Among these modes, $L_3$ is taken as the representative Laguerre--Gaussian case in the figures, while $L_4$ is retained only as a quantitative benchmark for radial-order dependence. This choice keeps the discussion focused because both modes exhibit the same qualitative ordering with respect to $\lambda$, whereas the differences between them are mainly quantitative.

For the visualizations discussed below, negative values of $\tilde\rho$ in the radial-cut and $(\tilde\rho,\tilde z)$ plots are shown only as a mirror representation of the axisymmetric profile about the origin; the physical radial coordinate satisfies $\tilde\rho \ge 0$.

\begin{figure}[H]
    \centering
    \begin{minipage}[b]{0.32\textwidth}
        \centering
        \includegraphics[width=\textwidth]{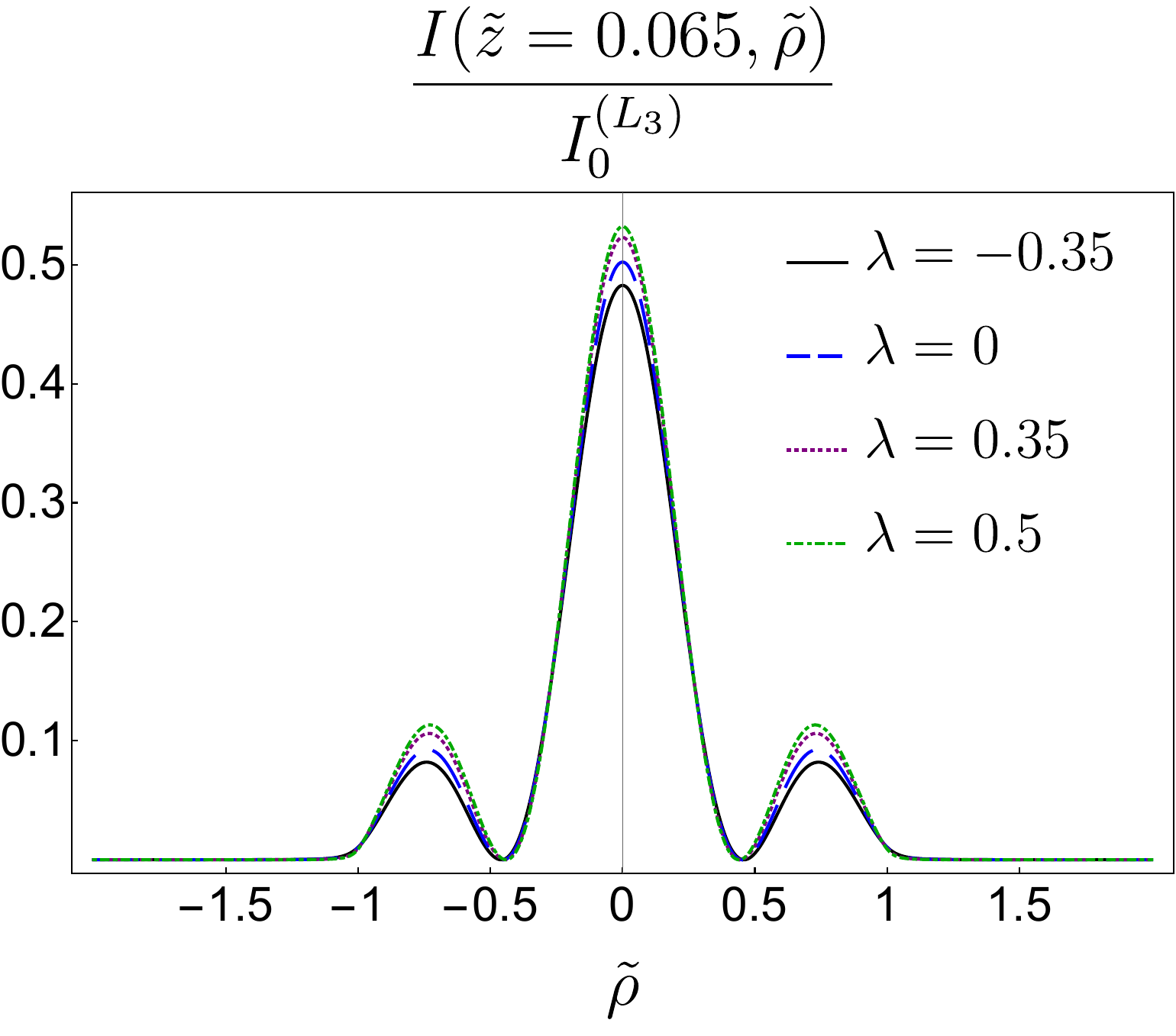}
        
        {\small (a) $\tilde z=0.065$}
    \end{minipage}
    \hfill
    \begin{minipage}[b]{0.328\textwidth}
        \centering
        \includegraphics[width=\textwidth]{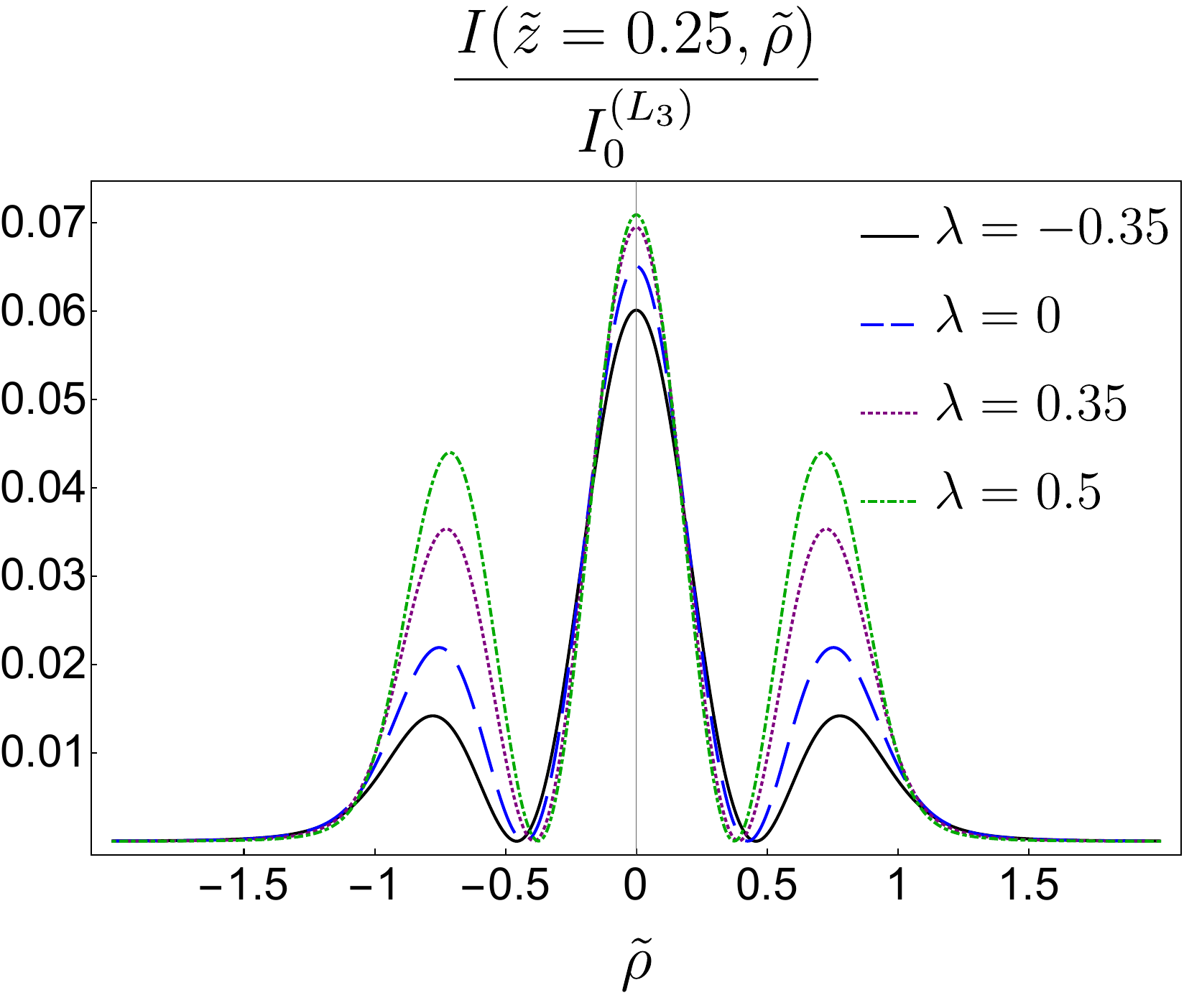}
        
        {\small (b) $\tilde z=0.25$}
    \end{minipage}
    \hfill
    \begin{minipage}[b]{0.337\textwidth}
        \centering
        \includegraphics[width=\textwidth]{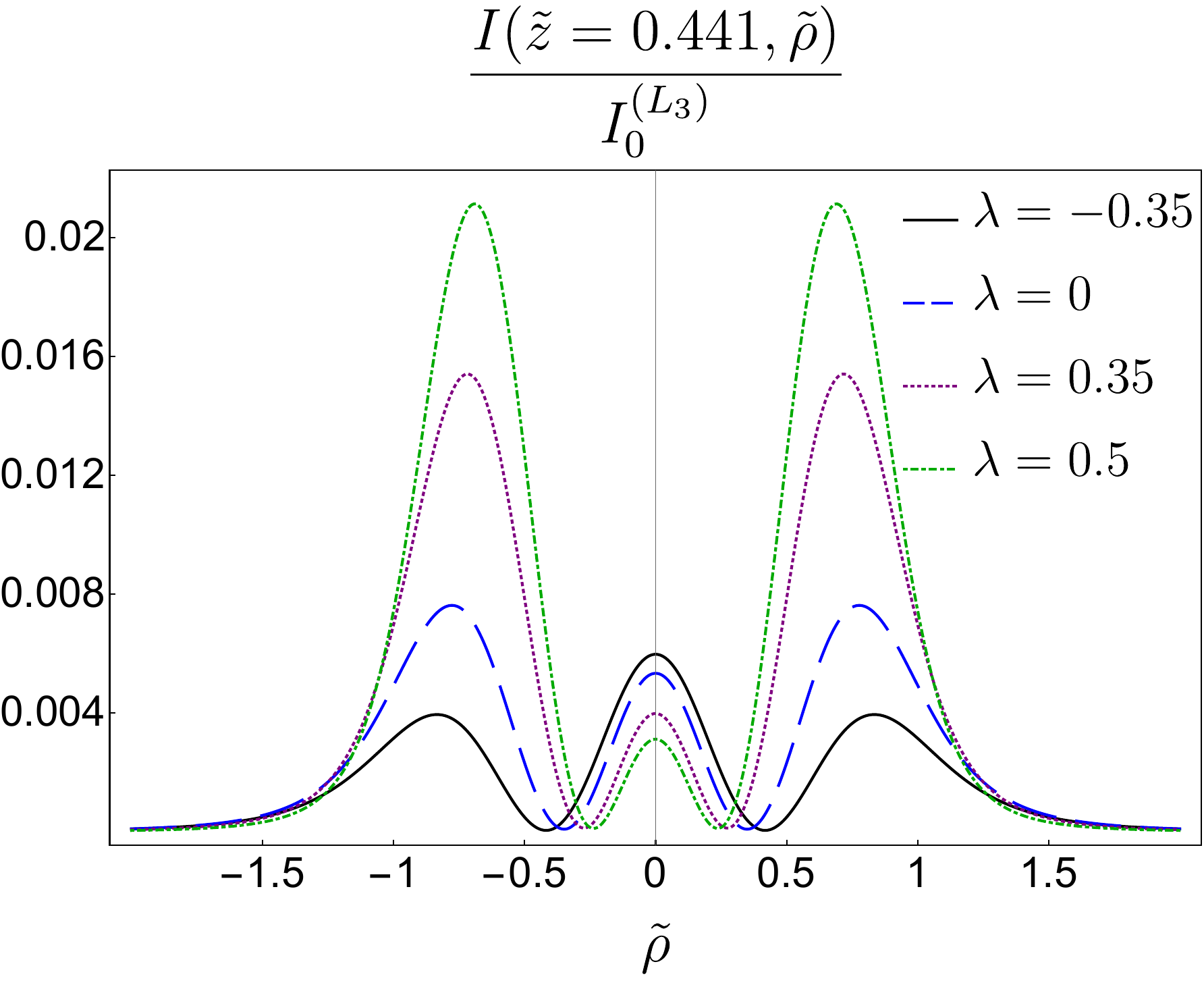}
        
        {\small (c) $\tilde z=0.441$}
    \end{minipage}
\caption{Radial intensity profiles $I(\tilde\rho,\tilde z)/I_0^{(L_3)}$ for the $L_3$ input at the three propagation distances used throughout the analysis. In each panel the four curves correspond to $\lambda=-0.35$, $0$, $0.35$, and $0.5$. All curves are normalized by the same entrance-plane reference intensity $I_0^{(L_3)} \equiv \max_{\tilde\rho} |\Phi_{\mathrm{in}}^{L_3}(\tilde\rho)|^2$.}
    \label{fig:L3profiles}
\end{figure}

The radial cuts in Fig.~\ref{fig:L3profiles} show a clear reorganization of the LG intensity with propagation. In the underlying $L_3$ datasets, the earliest sampled plane, $\tilde z=0.0025$, still lies very close to the prescribed entrance profile, with an on-axis normalized intensity within a few percent of unity for all values of $\lambda$. By $\tilde z=0.065$, the central intensity has already decreased markedly, although the dominant peak remains on axis for all four values of $\lambda$; its value increases monotonically from $I_{\rm peak}=0.4826$ at $\lambda=-0.35$ to $0.5321$ at $\lambda=0.5$. At the same time, the first radial minimum shifts slightly inward from $\tilde\rho_{\rm min}=0.460$ to $0.440$, indicating a mild compression of the central lobe. At $\tilde z=0.25$ the same ordering persists, with $I_{\rm peak}$ increasing from $0.0601$ to $0.0709$ as $\lambda$ grows.

The longer-distance cut at $\tilde z=0.441$ reveals the qualitative crossover that drives the later-stage dynamics. For $\lambda=-0.35$ the dominant maximum remains at the origin, with $I_{\rm peak}=I_{\rm center}=0.0060$. By contrast, for $\lambda=0$ the dominant maximum moves to $\tilde\rho_{\rm peak}=0.775$, where $I_{\rm peak}=0.0076$ while the on-axis value drops to $0.0053$. The same trend becomes stronger for $\lambda=0.35$ and $0.5$, where the dominant off-axis maxima move to $\tilde\rho_{\rm peak}=0.720$ and $0.690$ and the corresponding peak intensities rise to $0.0154$ and $0.0211$, respectively. In other words, positive anisotropy suppresses the central lobe at larger propagation distances and transfers the dominant weight toward side maxima at smaller radii. These trends are summarized quantitatively in Table~\ref{tab:L3obs}, which collects the selected observables for the three representative propagation distances. In particular, the table makes explicit the monotonic increase of the on-axis peak intensity with $\lambda$ at $\tilde z=0.065$ and $\tilde z=0.25$, as well as the transition at $\tilde z=0.441$ from an on-axis dominant maximum for $\lambda=-0.35$ to off-axis dominant maxima at progressively smaller radii as $\lambda$ increases. The axial summary in Table~\ref{tab:axialsummary} makes this ordering explicit: the first off-axis takeover occurs at $\tilde z_\times\approx 0.4917$, $0.4083$, $0.3417$, and $0.3083$ for $\lambda=-0.35$, $0$, $0.35$, and $0.5$, respectively. The monotonic decrease of $\tilde z_\times$ with $\lambda$ quantifies the progressive advance of off-axis dominance as the radial anisotropy strengthens. The $L_4$ benchmark follows the same ordering but with systematically earlier crossover values, which shows that increasing radial order amplifies the same reshaping mechanism rather than opening a distinct propagation regime.

\begin{table}[t]
\centering
\caption{Selected observables for the $L_3$ mode at the three propagation distances emphasized in the text.}
\label{tab:L3obs}
\begin{tabular}{ccccccc}
\toprule
$\tilde z$ & $\lambda$ & $\tilde\rho_{\rm peak}$ & $I_{\rm peak}$ & $I_{\rm center}$ & $r_{\rm eff}$ & $\tilde\rho_{\rm min}$ \\
\midrule
0.0650 & -0.35 & 0.000 & 0.4826 & 0.4826 & 0.618 & 0.460 \\
0.0650 & 0.00 & 0.000 & 0.5022 & 0.5022 & 0.624 & 0.450 \\
0.0650 & 0.35 & 0.000 & 0.5229 & 0.5229 & 0.629 & 0.445 \\
0.0650 & 0.50 & 0.000 & 0.5321 & 0.5321 & 0.632 & 0.440 \\
0.2500 & -0.35 & 0.000 & 0.0601 & 0.0601 & 0.817 & 0.455 \\
0.2500 & 0.00 & 0.000 & 0.0651 & 0.0651 & 0.810 & 0.425 \\
0.2500 & 0.35 & 0.000 & 0.0695 & 0.0695 & 0.779 & 0.390 \\
0.2500 & 0.50 & 0.000 & 0.0709 & 0.0709 & 0.760 & 0.375 \\
0.441 & -0.35 & 0.000 & 0.0060 & 0.0060 & 1.147 & 0.420 \\
0.441 & 0.00 & 0.775 & 0.0076 & 0.0053 & 1.062 & 0.350 \\
0.441 & 0.35 & 0.720 & 0.0154 & 0.0040 & 0.925 & 0.275 \\
0.441 & 0.50 & 0.690 & 0.0211 & 0.0031 & 0.854 & 0.235 \\
\bottomrule
\end{tabular}
\end{table}

\begin{figure}[H]
    \centering
    \begin{minipage}[t]{0.29\textwidth}
        \centering
        \vspace{0pt}
        \includegraphics[width=\textwidth]{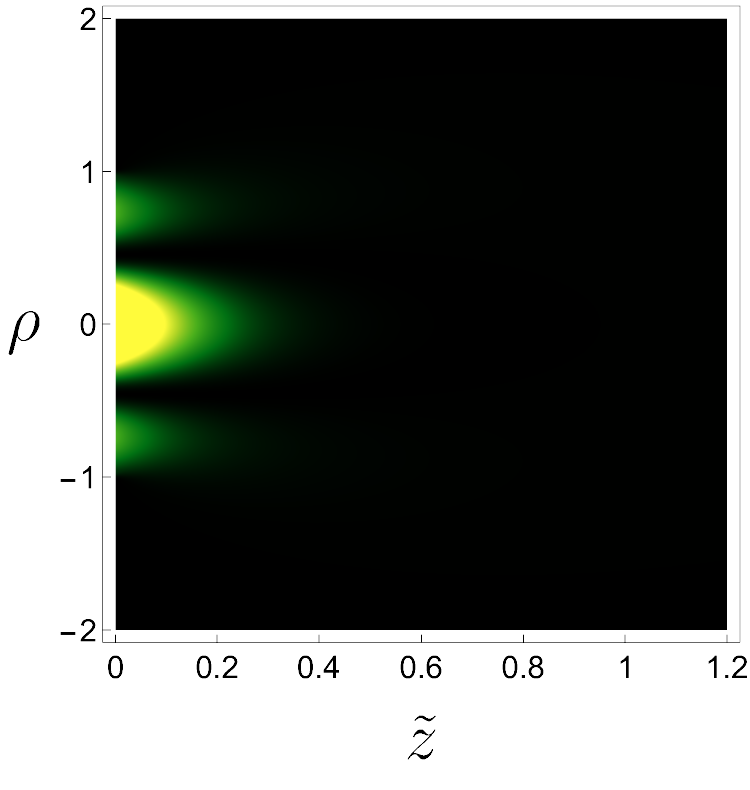}
        
        {\small (a) $\lambda=-0.35$}
    \end{minipage}
    \hfill
    \begin{minipage}[t]{0.29\textwidth}
        \centering
        \vspace{0pt}
        \includegraphics[width=\textwidth]{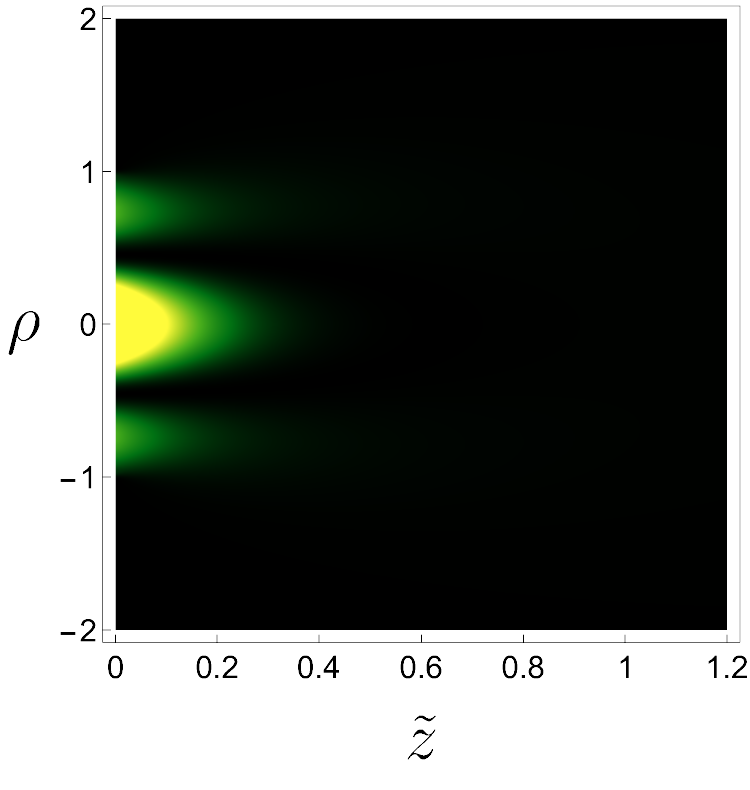}
        
        {\small (b) $\lambda=0$}
    \end{minipage}
    \hfill
    \begin{minipage}[t]{0.29\textwidth}
        \centering
        \vspace{0pt}
        \includegraphics[width=\textwidth]{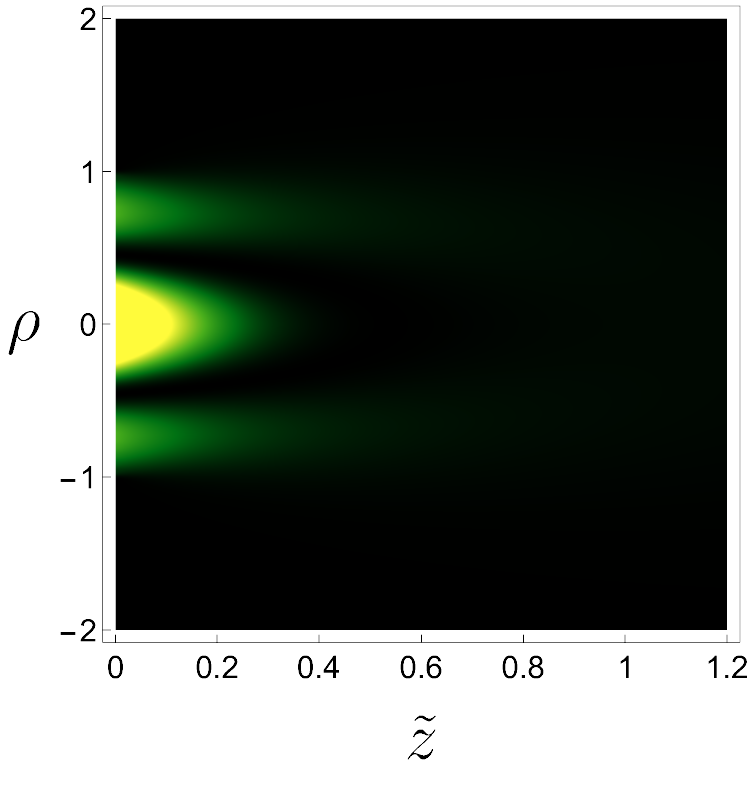}
        
        {\small (c) $\lambda=0.5$}
    \end{minipage}
    \hfill
    \begin{minipage}[t]{0.075\textwidth}
        \centering
        \vspace{-0.7cm}
        \includegraphics[width=\textwidth]{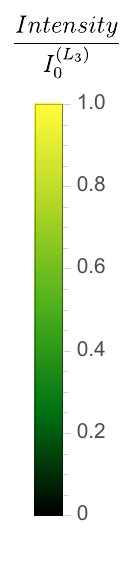}
    \end{minipage}
\caption{Maps of $I(\tilde\rho,\tilde z)/I_0^{(L_3)}$ for the $L_3$ input and three representative anisotropy values, $\lambda=-0.35$, $0$, and $0.5$. These maps summarize the continuous evolution behind the radial cuts of Fig.~1. All panels are normalized by the same entrance-plane reference intensity $I_0^{(L_3)} \equiv \max_{\tilde\rho} |\Phi_{\mathrm{in}}^{L_3}(\tilde\rho)|^2$.}
    \label{fig:L3density}
\end{figure}

The $(\tilde\rho,\tilde z)$ maps in Fig. \ref{fig:L3density} make the radial redistribution easier to read globally. For $\lambda=-0.35$, the on-axis maximum remains dominant over a longer interval, and the first side maximum develops more slowly. For $\lambda=0.5$, the side maxima emerge earlier and subsequently overtake the central peak. Thus the effect of increasing $\lambda$ is not a simple longitudinal translation of the whole pattern but a genuine reweighting of radial channels.

\begin{figure}[H]
    \centering
    \begin{minipage}[t]{0.29\textwidth}
        \centering
        \vspace{0pt}
        \includegraphics[width=\textwidth]{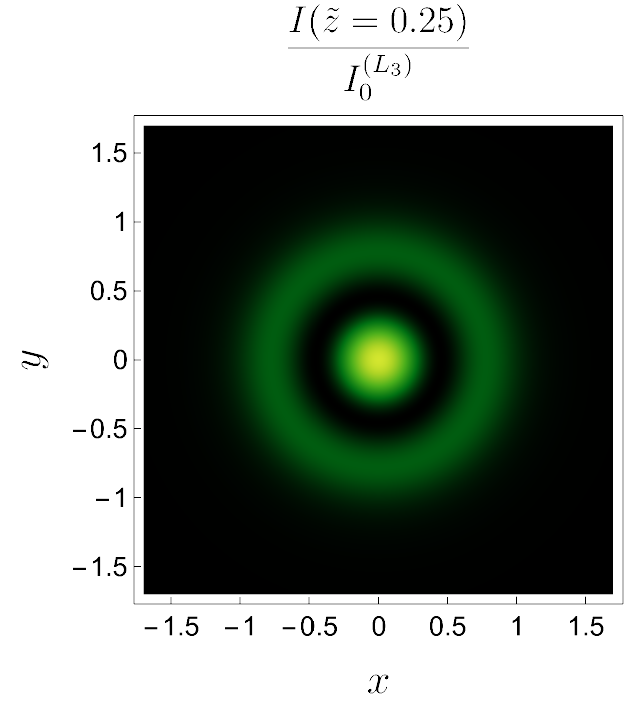}
        
        {\small (a) $\lambda=-0.35$}
    \end{minipage}
    \hfill
    \begin{minipage}[t]{0.29\textwidth}
        \centering
        \vspace{0pt}
        \includegraphics[width=\textwidth]{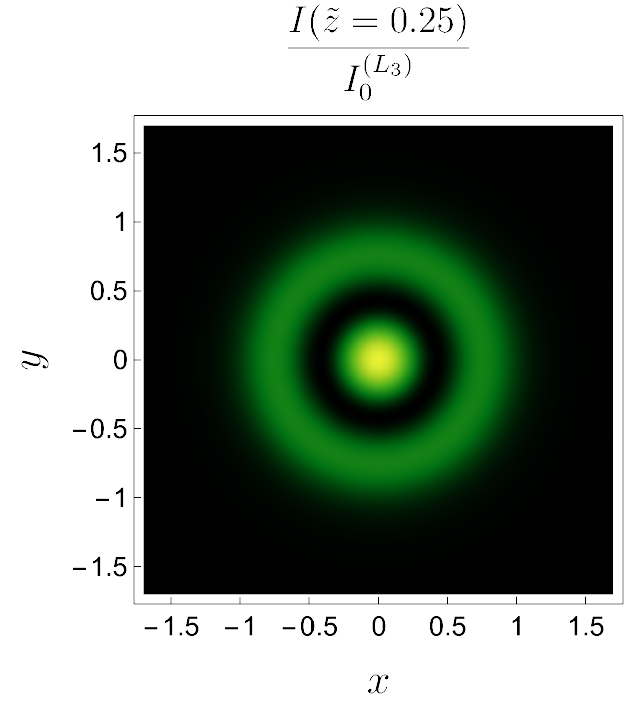}
        
        {\small (b) $\lambda=0$}
    \end{minipage}
    \hfill
    \begin{minipage}[t]{0.29\textwidth}
        \centering
        \vspace{0pt}
        \includegraphics[width=\textwidth]{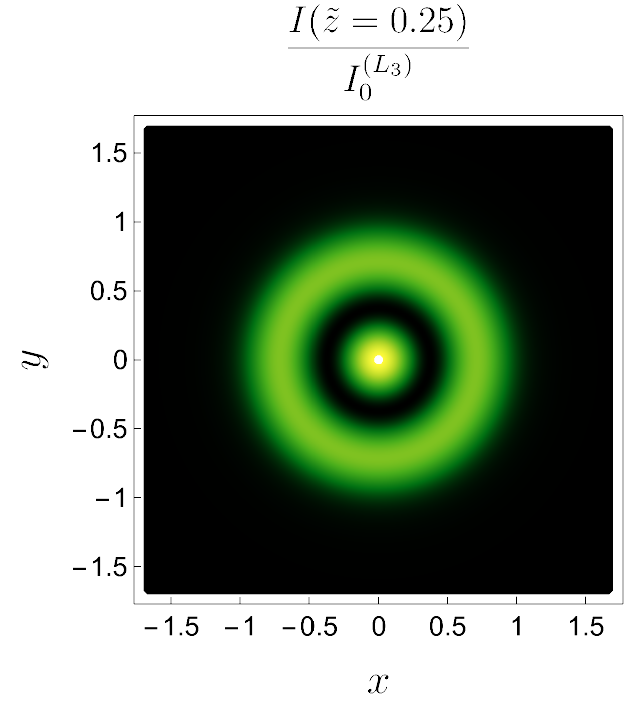}
        
        {\small (c) $\lambda=0.5$}
    \end{minipage}
    \hfill
    \begin{minipage}[t]{0.08\textwidth}
        \centering
        \vspace{-0.5cm}
        \includegraphics[width=\textwidth]{BarraL3.pdf}
    \end{minipage}
\caption{Transverse intensity maps $I(x,y;\tilde z)/I_0^{(L_3)}$ for the $L_3$ mode at $\tilde z=0.25$ and $\lambda=-0.35$, $0$, and $0.5$. The azimuthal symmetry is preserved, while the ring contrast and effective radial scale depend strongly on $\lambda$. All panels are normalized by the same entrance-plane reference intensity $I_0^{(L_3)} \equiv \max_{\tilde\rho} |\Phi_{\mathrm{in}}^{L_3}(\tilde\rho)|^2$.}
    \label{fig:L3rings}
\end{figure}

The transverse patterns in Fig.~\ref{fig:L3rings} provide the same information in a form closer to the beam images encountered in structured-light experiments. Negative $\lambda$ produces a broader central spot with less pronounced side maxima, whereas positive $\lambda$ yields sharper ring contrast and stronger off-axis localization. Because the configuration is axially symmetric, these changes are entirely radial and do not rely on angular-mode mixing. This should be contrasted with propagation in uniaxial crystals or bianisotropic settings, where polarization conversion, walk-off, and ordinary/extraordinary splitting also participate in the reshaping process \cite{Ciattoni2002Cylindrical,Cincotti2002Uniaxial,Egorov2024Uniaxial,Berskys2022SphericalVBessel}. The present model isolates only the scalar radial channel, so the comparison with that literature is qualitative rather than one-to-one. In that limited but controlled sense, the present LG results identify the part of the reshaping that can already be understood as radial reweighting before polarization-dependent mechanisms are restored.

Taken together, these results show that for $L_3$ modes negative $\lambda$ delays the off-axis takeover and preserves the central lobe over a longer propagation interval, whereas positive $\lambda$ advances $\tilde z_\times$ and drives a stronger transfer of intensity toward side maxima.

\subsection{Quantitative $L_4$ benchmark}

The $L_4$ mode follows the same qualitative $\lambda$-dependent trends as $L_3$ and therefore does not require a second full set of figures. Its main role is to quantify how the radial order modifies the same underlying mechanism. Table~\ref{tab:L4vsL3} shows that $L_4$ is systematically more compact than $L_3$ and exhibits lower dominant-peak intensity at all three representative propagation distances. Near the entrance plane, the peak-intensity ratio $I^{(L_4)}_{\rm peak}/I^{(L_3)}_{\rm peak}$ is approximately $0.91$ for all $\lambda$, while the effective-radius ratio is about $0.88$. At $\tilde z=0.25$ the peak-intensity ratio decreases to $0.67$--$0.68$ and the radius ratio remains between $0.89$ and $0.92$. At $\tilde z=0.441$ the same qualitative behavior persists, with the peak-intensity ratio spanning $0.47$--$0.73$. The strongest off-axis maximum also appears at slightly smaller radius for $L_4$ than for $L_3$ once the side lobes dominate. Consistently with Table~\ref{tab:axialsummary}, the onset of off-axis dominance occurs earlier for $L_4$ than for $L_3$ at every $\lambda$, which quantifies the greater sensitivity of higher radial order without altering the qualitative ordering of the anisotropy effect. Thus $L_4$ reinforces the same anisotropy trends observed in $L_3$ without introducing a qualitatively new regime, which is why we do not include a second complete visual sequence for it.

\begin{table}[t]
\centering
\caption{Quantitative comparison between the $L_3$ and $L_4$ modes at the representative propagation distances.}
\label{tab:L4vsL3}
\begin{tabular}{cccccccc}
\toprule
$\tilde z$ & $\lambda$ & $I^{(L_4)}_{\rm peak}$ & $I^{(L_3)}_{\rm peak}$ & $I^{(L_4)}_{\rm peak}/I^{(L_3)}_{\rm peak}$ & $r^{(L_4)}_{\rm eff}$ & $r^{(L_3)}_{\rm eff}$ & $r^{(L_4)}_{\rm eff}/r^{(L_3)}_{\rm eff}$ \\
\midrule
0.0650 & -0.35 & 0.4369 & 0.4826 & 0.905 & 0.543 & 0.618 & 0.878 \\
0.0650 & 0.00 & 0.4568 & 0.5022 & 0.910 & 0.548 & 0.624 & 0.879 \\
0.0650 & 0.35 & 0.4781 & 0.5229 & 0.914 & 0.554 & 0.629 & 0.880 \\
0.0650 & 0.50 & 0.4874 & 0.5321 & 0.916 & 0.556 & 0.632 & 0.881 \\
0.2500 & -0.35 & 0.0410 & 0.0601 & 0.681 & 0.750 & 0.817 & 0.918 \\
0.2500 & 0.00 & 0.0441 & 0.0651 & 0.678 & 0.738 & 0.810 & 0.911 \\
0.2500 & 0.35 & 0.0468 & 0.0695 & 0.673 & 0.699 & 0.779 & 0.897 \\
0.2500 & 0.50 & 0.0475 & 0.0709 & 0.670 & 0.676 & 0.760 & 0.890 \\
0.441 & -0.35 & 0.0028 & 0.0060 & 0.468 & 1.096 & 1.147 & 0.955 \\
0.441 & 0.00 & 0.0052 & 0.0076 & 0.684 & 0.993 & 1.062 & 0.935 \\
0.441 & 0.35 & 0.0111 & 0.0154 & 0.718 & 0.837 & 0.925 & 0.905 \\
0.441 & 0.50 & 0.0154 & 0.0211 & 0.730 & 0.759 & 0.854 & 0.889 \\
\bottomrule
\end{tabular}
\end{table}

\subsection{Bessel--Gaussian beams}

Bessel--Gaussian beams complement the Laguerre family because they combine a ringed Bessel-type core with a Gaussian envelope that renders the field finite in energy and experimentally accessible. This makes them especially useful for asking how an effective radial anisotropy acts on a beam that is known to preserve its transverse structure over extended---but not infinite---distances, and for distinguishing genuine radial redistribution from the ordinary broadening of a finite-aperture input. The choice $\tilde k_r=5.9$ used below places both the bright core and the first ring well inside the normalized aperture window, which makes the evolution of the central lobe and of the surrounding ring straightforward to quantify.

The BG input used in this work is
\begin{equation}
\Phi_{\mathrm{in}}^{\mathrm{BG}}(\tilde\rho)=J_0(\tilde k_r\tilde\rho)e^{-\kappa^2\tilde\rho^2}\,\Theta(1-\tilde\rho),
\label{eq:BGinput}
\end{equation}

Finite-energy Bessel--Gaussian beams inherit the familiar long-depth, self-healing intuition of ideal Bessel beams, but their Gaussian envelope and finite aperture prevent exact diffractionless propagation \cite{Durnin1987,Durnin1987PRL,Khonina2020}. The physically relevant question is therefore how the approximately diffraction-resistant ring structure survives as $\lambda$ varies. 

Although Table~\ref{tab:z0validation} focuses on the truncated Laguerre inputs, the BG launch-plane profiles are also numerically well controlled: the maximum pairwise relative $L^2$ difference among distinct $\lambda$ values at $\tilde z=0$ remains below $1.1\times10^{-2}$. This does not replace the reconstruction metrics reported for LG, but it does show that the BG differences discussed below are not driven by mismatched initial profiles.

\begin{figure}[H]
    \centering
    \begin{minipage}[b]{0.32\textwidth}
        \centering
        \includegraphics[width=\textwidth]{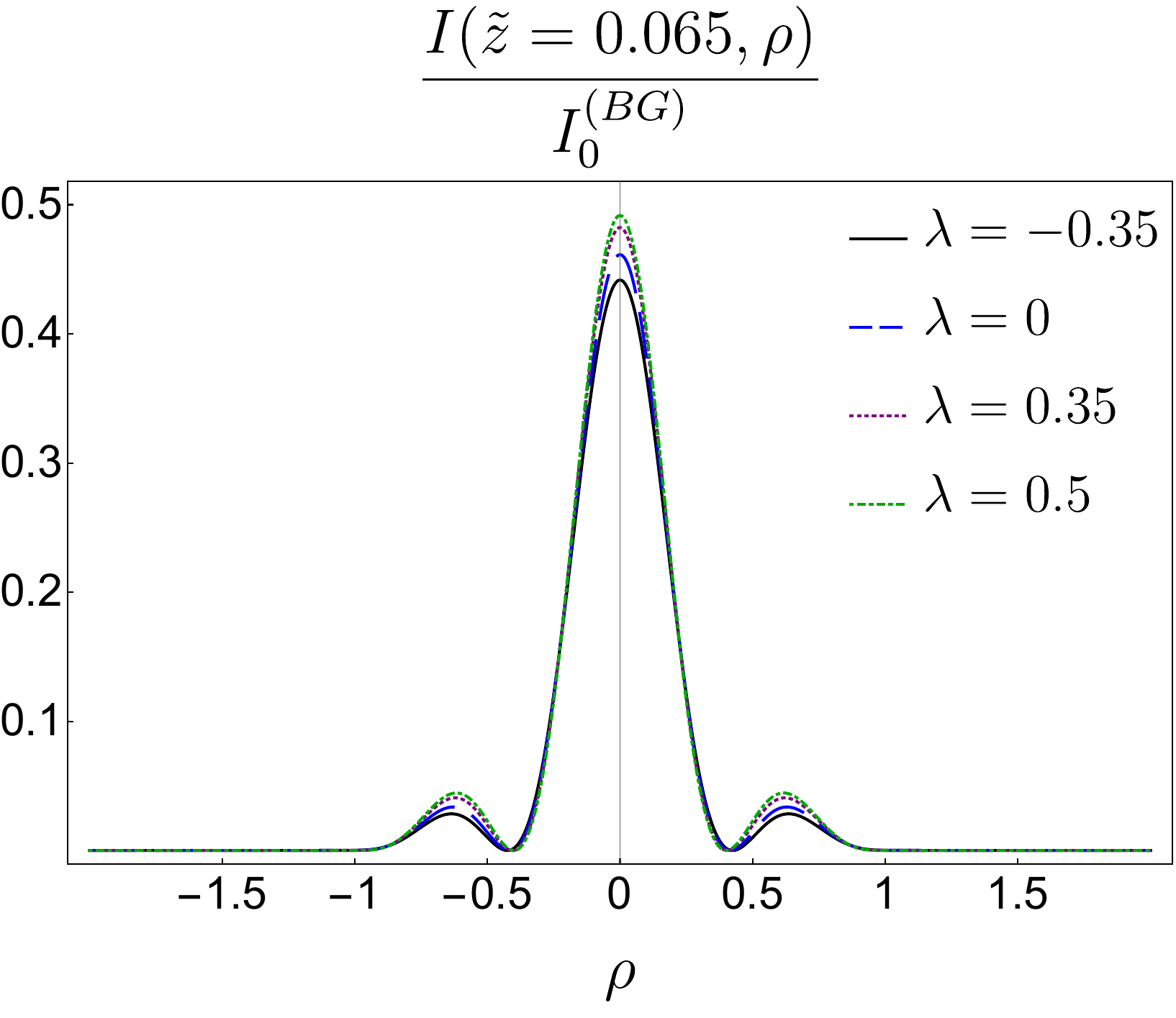}
        
        {\small (a) $\tilde z=0.065$}
    \end{minipage}
    \hfill
    \begin{minipage}[b]{0.328\textwidth}
        \centering
        \includegraphics[width=\textwidth]{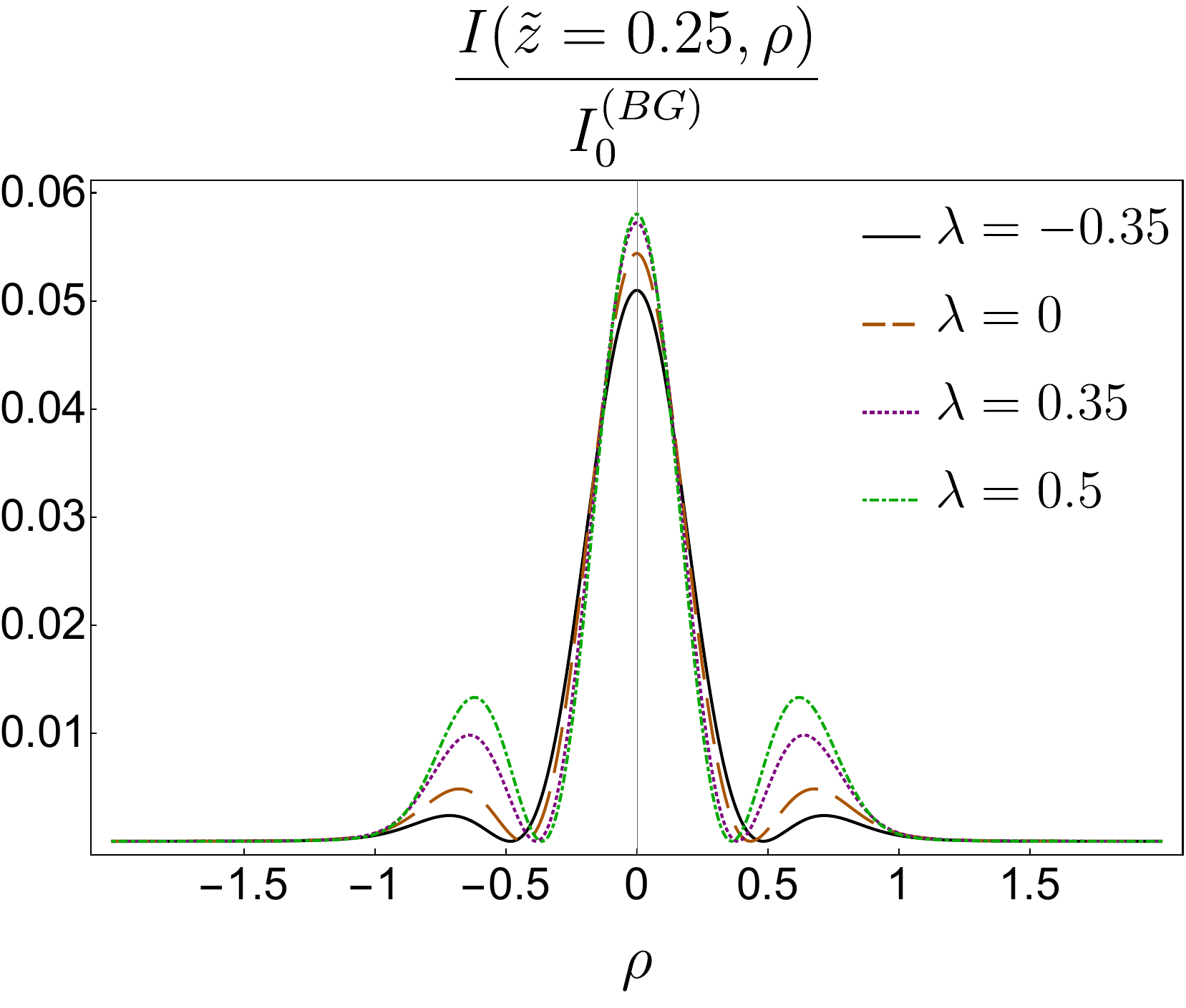}
        
        {\small (b) $\tilde z=0.25$}
    \end{minipage}
    \hfill
    \begin{minipage}[b]{0.337\textwidth}
        \centering
        \includegraphics[width=\textwidth]{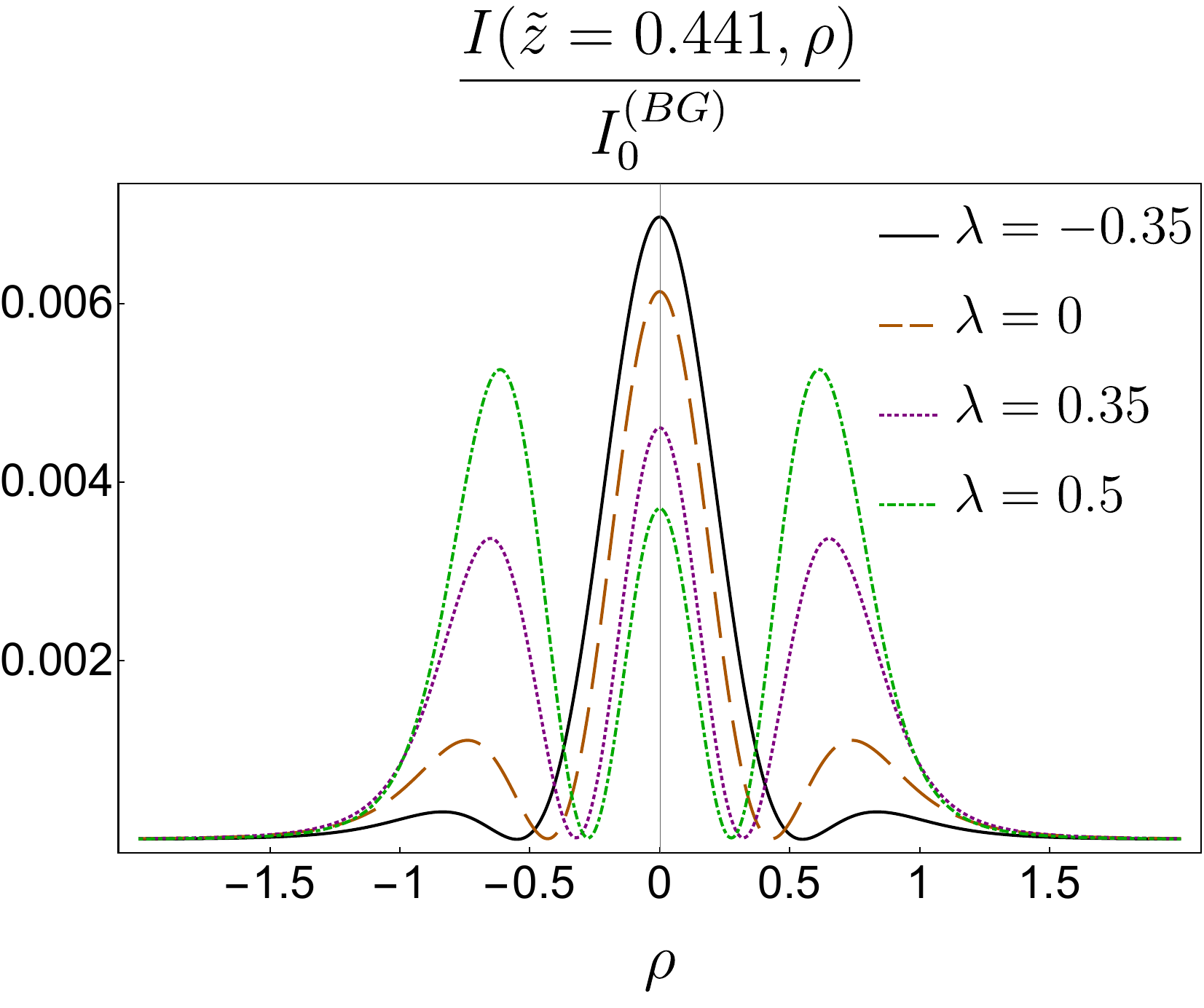}
        
        {\small (c) $\tilde z=0.441$}
    \end{minipage}
    \caption{Radial intensity profiles $I(\tilde\rho,\tilde z)/I_0^{(\mathrm{BG})}$ for the BG input at $\tilde z=0.065$, $0.25$, and $0.441$, with $\lambda=-0.35$, $0$, $0.35$, and $0.5$. All curves are normalized by the same entrance-plane reference intensity $I_0^{(\mathrm{BG})} \equiv \max_{\tilde\rho} |\Phi_{\mathrm{in}}^{\mathrm{BG}}(\tilde\rho)|^2$.}
    \label{fig:BGprofiles}
\end{figure}

\begin{figure}[H]
    \centering
    \begin{minipage}[t]{0.29\textwidth}
        \centering
        \vspace{0pt}
        \includegraphics[width=\textwidth]{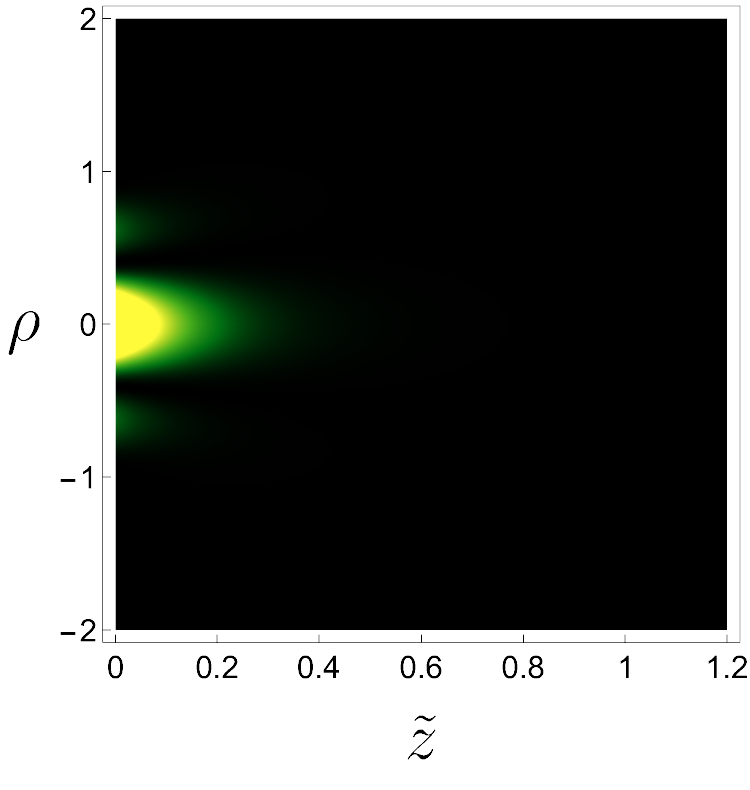}
        
        {\small (a) $\lambda=-0.35$}
    \end{minipage}
    \hfill
    \begin{minipage}[t]{0.29\textwidth}
        \centering
        \vspace{0pt}
        \includegraphics[width=\textwidth]{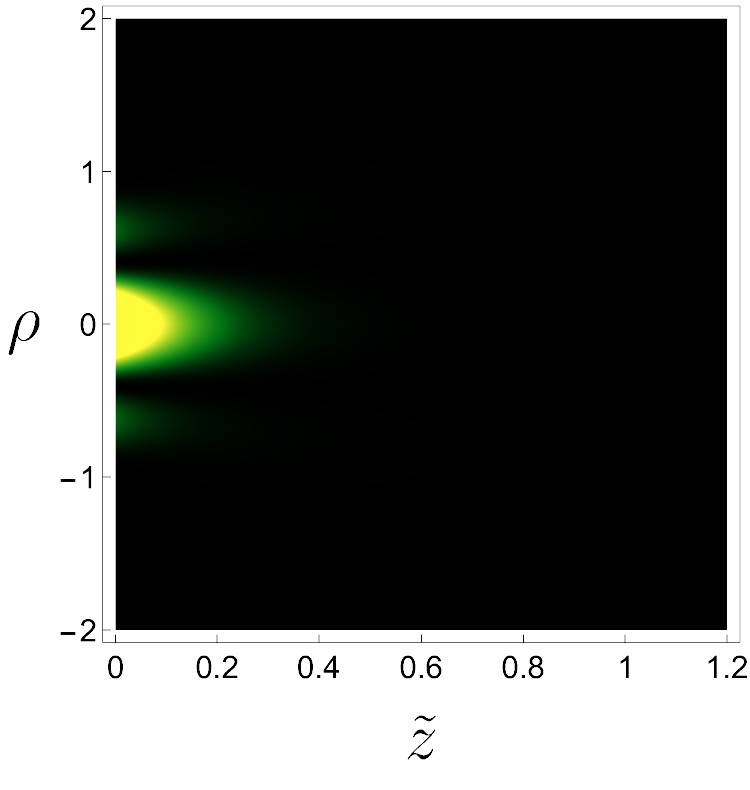}
        
        {\small (b) $\lambda=0$}
    \end{minipage}
    \hfill
    \begin{minipage}[t]{0.29\textwidth}
        \centering
        \vspace{0pt}
        \includegraphics[width=\textwidth]{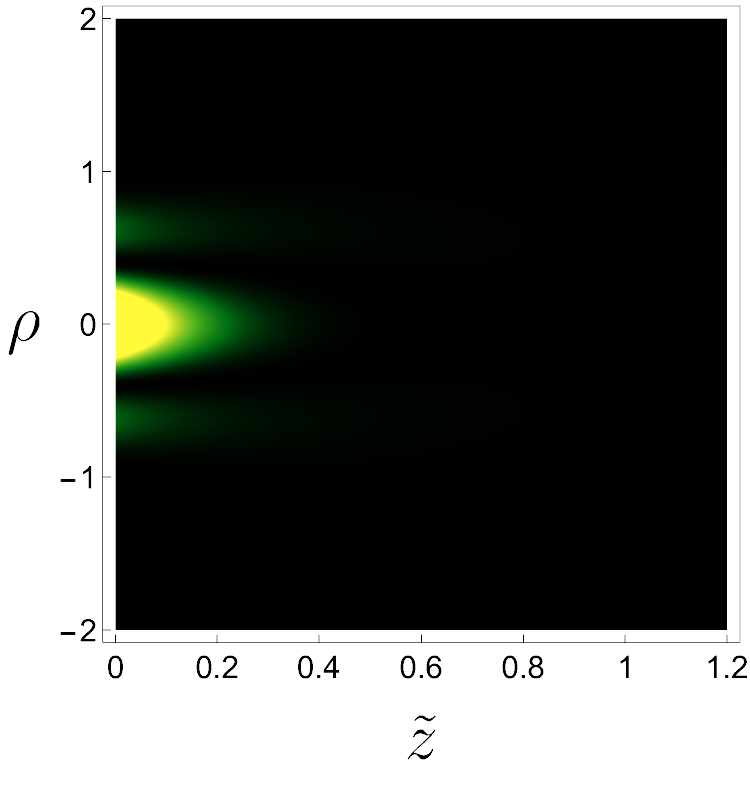}
        
        {\small (c) $\lambda=0.5$}
    \end{minipage}
    \hfill
    \begin{minipage}[t]{0.075\textwidth}
        \centering
        \vspace{-0.7cm}
        \includegraphics[width=\textwidth]{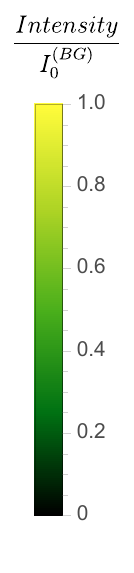}
    \end{minipage}
 \caption{Maps of $I(\tilde\rho,\tilde z)/I_0^{(\mathrm{BG})}$ for the BG input and three representative anisotropy values, $\lambda=-0.35$, $0$, and $0.5$. These maps summarize the continuous evolution behind the radial cuts of Fig.~4. All panels are normalized by the same entrance-plane reference intensity $I_0^{(\mathrm{BG})} \equiv \max_{\tilde\rho} |\Phi_{\mathrm{in}}^{\mathrm{BG}}(\tilde\rho)|^2$.}
    \label{fig:BGdensity}
\end{figure}

\begin{figure}[H]
    \centering
    \begin{minipage}[t]{0.29\textwidth}
        \centering
        \vspace{0pt}
        \includegraphics[width=\textwidth]{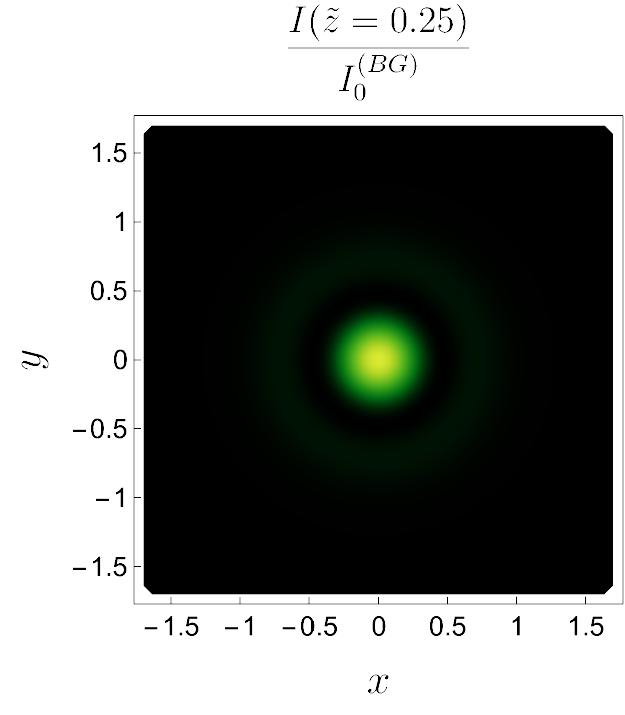}
        
        {\small (a) $\lambda=-0.35$}
    \end{minipage}
    \hfill
    \begin{minipage}[t]{0.29\textwidth}
        \centering
        \vspace{0pt}
        \includegraphics[width=\textwidth]{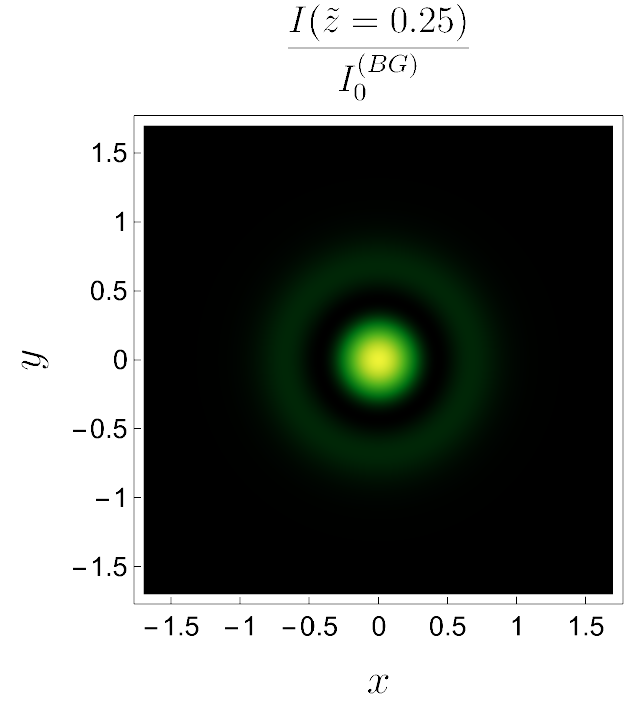}
        
        {\small (b) $\lambda=0$}
    \end{minipage}
    \hfill
    \begin{minipage}[t]{0.29\textwidth}
        \centering
        \vspace{0pt}
        \includegraphics[width=\textwidth]{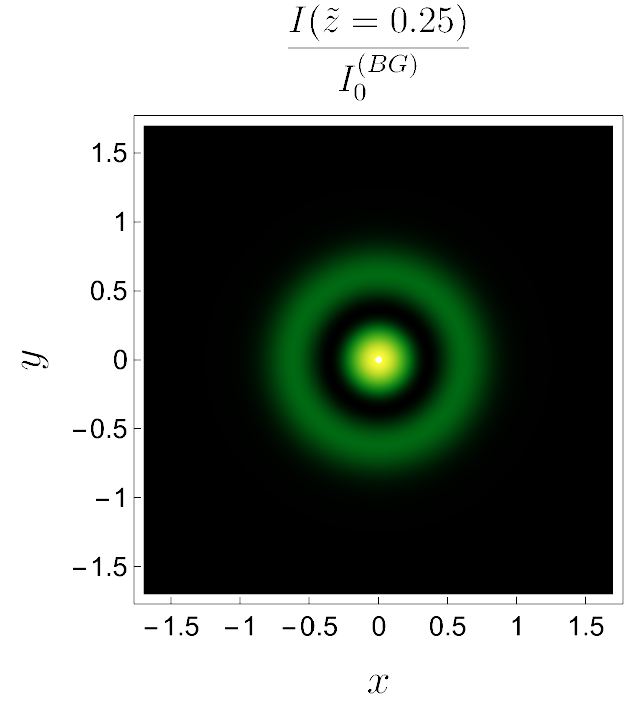}
        
        {\small (c) $\lambda=0.5$}
    \end{minipage}
    \hfill
    \begin{minipage}[t]{0.08\textwidth}
        \centering
        \vspace{-0.5cm}
        \includegraphics[width=\textwidth]{BarraBG.pdf}
    \end{minipage}
\caption{Transverse intensity maps $I(x,y;\tilde z)/I_0^{(\mathrm{BG})}$ for the BG mode at $\tilde z=0.25$ and $\lambda=-0.35$, $0$, and $0.5$. All panels are normalized by the same entrance-plane reference intensity $I_0^{(\mathrm{BG})} \equiv \max_{\tilde\rho} |\Phi_{\mathrm{in}}^{\mathrm{BG}}(\tilde\rho)|^2$.}
    \label{fig:BGrings}
\end{figure}

From the radial cuts in Fig.~\ref{fig:BGprofiles} we see that at $\tilde z=0.065$ the BG beam remains strongly on-axis for all values of $\lambda$, with $I_{\rm center}$ ranging from $0.4416$ at $\lambda=-0.35$ to $0.4916$ at $\lambda=0.5$. At $\tilde z=0.25$ the on-axis maximum persists, but the first radial minimum moves inward as $\lambda$ increases, from $\tilde\rho_{\rm min}=0.480$ to $0.360$. The largest contrast appears at $\tilde z=0.441$: for $\lambda\le 0.35$ the dominant maximum is still central, whereas for $\lambda=0.5$ the strongest maximum moves off axis to $\tilde\rho_{\rm peak}=0.615$, where $I_{\rm peak}=0.00526$ while the on-axis value is $0.00371$. Compared with the LG case, the BG family therefore retains its central structure for a longer propagation interval. This is consistent with the broader Bessel--Gaussian literature, where finite-energy and vector BG beams preserve their ring morphology over extended but finite distances rather than exhibiting exact nondiffracting behavior \cite{Durnin1987,Durnin1987PRL,Huang2011VectorVortexBG,Berskys2022SphericalVBessel}. Table~\ref{tab:axialsummary} makes the comparison with $L_3$ explicit: for $\lambda=0$, $0.35$, and $0.5$, the BG crossover distances $\tilde z_\times=0.6750$, $0.4750$, and $0.4167$ are all larger than their $L_3$ counterparts. For the BG beam with $\lambda=-0.35$, no off-axis crossover is observed within the propagation interval considered here. In other words, the on-axis maximum remains dominant throughout the explored window, so no finite value of $\tilde z_\times$ can be assigned and, correspondingly, there is no associated off-axis $\tilde\rho_{\mathrm{peak}}(\tilde z_\times)$ in that range. This is consistent with the broader profile and the delayed radial reorganization already seen in Figs.~\ref{fig:BGprofiles}--\ref{fig:BGrings} and supports the interpretation that negative $\lambda$ preserves the central BG structure over a longer propagation distance. In this sense BG inputs preserve a dominant central lobe over a longer interval, so the anisotropy first appears as a broadening or compression of the ringed core before it produces a genuine off-axis takeover.

The BG density maps in Fig.~\ref{fig:BGdensity} display the same trend continuously. Negative $\lambda$ maintains a broader on-axis channel over a longer propagation interval, whereas positive $\lambda$ accelerates the inward motion of the first minimum and amplifies the side rings. As in the LG case, the effect is best understood as a reweighting among radial channels rather than as a rigid axial shift of the entire pattern. Relative to vectorial BG propagation in anisotropic media, the present result should be read as the scalar-effective component of that reshaping: it tracks how much of the change can already be attributed to a radial anisotropy of the propagation operator before polarization mixing is restored. This is precisely where the present calculation complements the optical literature: it provides a baseline for the radial part of the reshaping, to which polarization-dependent anisotropic effects can later be added in more complete models.

The transverse BG patterns in Fig.~\ref{fig:BGrings} present the same information in a plane-wavefront image representation. Negative $\lambda$ broadens the bright core and weakens the surrounding ring, whereas positive $\lambda$ increases the ring contrast and contracts the overall radial scale. The azimuthal symmetry is again preserved, so the anisotropy acts through radial redistribution alone.

\begin{table}[t]
\centering
\caption{Quantitative indicators of BG shape preservation between $\tilde z=0$ and $\tilde z=0.441$.}
\label{tab:BGdiffraction}
\begin{tabular}{cccccc}
\toprule
$\lambda$ & $\mathrm{FWHM}(0.441)/\mathrm{FWHM}(0)$ & $r_{\rm eff}(0.441)/r_{\rm eff}(0)$ & $I_{\rm center}(0.441)/I_{\rm center}(0)$ & $\tilde\rho_{\rm min}(0)$ & $\tilde\rho_{\rm min}(0.441)$ \\
\midrule
-0.35 & 1.285 & 1.658 & 0.0070 & 0.410 & 0.550 \\
0.00  & 1.085 & 2.110 & 0.0062 & 0.410 & 0.430 \\
0.35  & 0.855 & 1.985 & 0.0047 & 0.410 & 0.320 \\
0.50  & 0.739 & 1.821 & 0.0039 & 0.410 & 0.275 \\
\bottomrule
\end{tabular}
\end{table}

A useful quantitative measure of the fate of the diffraction-resistant structure is provided by the evolution of the central-lobe width and of the first radial minimum. Table~\ref{tab:BGdiffraction} compares these quantities between $\tilde z=0$ and $\tilde z=0.441$. For $\lambda=-0.35$, the full width at half maximum (FWHM) grows by about $28\%$, indicating broadening of the central lobe. For $\lambda=0$, the increase is much milder ($\approx 8.5\%$). By contrast, for $\lambda=0.35$ and $0.5$ the FWHM decreases to roughly $85\%$ and $74\%$ of its entrance-plane value, respectively. Simultaneously, the first radial minimum shifts from $\tilde\rho\approx 0.41$ at launch to $0.55$, $0.43$, $0.32$, and $0.275$ as $\lambda$ runs from $-0.35$ to $0.5$. Thus the scalar anisotropy does not simply destroy the BG morphology; it either broadens or compresses the finite-energy ring structure depending on the sign and magnitude of $\lambda$.

\section{Conclusion}
\label{Conclu1}

We have presented a study of LG and BG scalar-beam propagation in an effective anisotropic background controlled by a single parameter $\lambda$. The model is intentionally scalar: it does not replace the full dielectric-tensor description of an anisotropic crystal and therefore does not address polarization-dependent birefringence. Instead, it isolates how a transverse radial director field modifies the radial part of the propagation operator.

The numerical analysis was carried out with an initial-value spectral
reconstruction formulated for the same axisymmetric propagation problem as
the Green-kernel description. Within the numerical accuracy quantified in Table~\ref{tab:z0validation}, the prescribed entrance profile is recovered at $\tilde z=0$ for every value of $\lambda$, which removes the spurious input-plane discrepancies and makes it possible to compare quantitative observables along propagation. The axial summaries in Table~\ref{tab:axialsummary} further show that the maximum on-axis intensity always occurs at launch and that the minimum effective radius is reached very close to the launch plane. For the Laguerre--Gaussian family, $L_3$ captures the representative dynamics: positive $\lambda$ enhances the transfer of weight from the central lobe to off-axis maxima at sufficiently large $\tilde z$, while negative $\lambda$ delays that crossover. The $L_4$ mode follows the same qualitative ordering with respect to $\lambda$ and therefore serves as a quantitative benchmark rather than a separate visual case.

For finite-aperture Bessel--Gaussian beams, the effective anisotropy modifies how the approximately diffraction-resistant ring structure evolves. Negative $\lambda$ broadens the central region and pushes the first minimum outward, while sufficiently positive $\lambda$ compresses the profile and eventually promotes an off-axis dominant maximum. The delayed crossover values of Table~\ref{tab:axialsummary} quantify that BG inputs preserve their central structure over a longer interval than LG inputs. These effects are naturally described as radial intensity redistribution induced by the effective anisotropic background.

The results support a clear interpretation of $\lambda$ as a reduced anisotropy parameter in a scalar-effective propagation problem: within the axially symmetric sector, it measures the relative weight of radial derivatives rather than a full constitutive response. The novelty is therefore not a crystal-specific constitutive description but a nonparaxial launch-plane analysis that isolates the radial scalar channel and shows how it reorganizes LG and BG beams in systematically different ways. Within that scope, structured beams provide a controlled setting in which radial reshaping, side-lobe amplification, delayed or accelerated off-axis takeover, and ring displacement can be quantified without invoking a full vectorial medium model. The comparison with vectorial crystal and bianisotropic studies is therefore complementary: those works resolve polarization-sensitive anisotropy, whereas the present one identifies the baseline radial contribution against which more complete optical descriptions and experiments can be interpreted.

Natural extensions include genuinely vortex-carrying modes with $m\neq 0$, vectorial generalizations that incorporate polarization, and explicit analogue implementations in engineered anisotropic platforms.

\begin{acknowledgments}
R. L. acknowledges partial support from CONAHCyT-M\'exico under Grant No. CBF-2023-2024-1937.
E. P. F. thanks Secretar\'ia de Ciencias, Humanidades, Tecnolog\'ia e Innovaci\'on (SECIHTI, M\'exico) and Universidad Aut\'onoma Metropolitana--Iztapalapa for continued support.
\end{acknowledgments}

\section*{Data availability}
Numerical datasets underlying the figures and tables were generated from the spectral reconstruction described in Sec.~\ref{Model1} and are available from the corresponding author upon reasonable request.

\appendix

\section{From the Green kernel to the initial-value spectral representation}
\label{App1}

This appendix summarizes the steps that connect the axisymmetric Green-kernel representation with the spectral formulas used in the numerical implementation. Related Fourier--Hankel spectral representations are widely used for the propagation of axisymmetric scalar optical fields and for the numerical implementation of cylindrical propagation operators \cite{Yu1998QDHT,GuizarSicairos2004QDHT,Kirilenko2018Axisymmetric}; the present construction follows the same general logic, although here the anisotropic operator induces the modified radial weight $\tilde\rho^\sigma$ and the branch-adapted basis $J_\lambda$. We work throughout in the $m=0$ sector and in the launch-plane description at $z'=0$.

Starting from the Green kernel in Eq.~\eqref{eq:greenKernel}, and suppressing the trivial angular dependence, the launch-plane representation reads
\begin{equation}
G_0(\rho,z;\rho',0)
=
\frac{i\pi}{1-\lambda}
\left(\frac{\rho}{\rho'}\right)^{\sigma}
\int_{-\infty}^{\infty}\frac{dk}{(2\pi)^2}\,
e^{ik z}\,
J_{\nu_0}(\chi \rho_<)\,
H_{\nu_0}^{(1)}(\chi \rho_>),
\label{eq:App_green_m0}
\end{equation}
where
\begin{equation}
\sigma=\frac{\lambda}{2(\lambda-1)},\qquad
\nu_0=\frac{|\lambda|}{2(1-\lambda)},\qquad
\chi=\frac{\sqrt{(\omega/c)^2-k^2}}{\sqrt{1-\lambda}},
\label{eq:App_green_params}
\end{equation}
and $\rho_< = \min(\rho,\rho')$, $\rho_> = \max(\rho,\rho')$. In this form, the factor
\(
J_{\nu_0}(\chi \rho_<)H_{\nu_0}^{(1)}(\chi \rho_>)
\)
encodes the standard regular--outgoing decomposition of the radial propagator.

For the numerical problem studied in the main text, however, the field is specified by a prescribed entrance profile at the launch plane rather than by an arbitrary bulk source. It is therefore convenient to solve the corresponding homogeneous propagation problem and determine the expansion coefficients directly from the launch data.

In dimensionless variables,
\begin{equation}
\tilde\rho=\rho/\rho_0,\qquad
\tilde z=z/\rho_0,\qquad
\Omega=\omega\rho_0/c,
\label{eq:App_dimensionless}
\end{equation}
the axisymmetric homogeneous equation reads
\begin{equation}
\left[
\frac{\partial^2}{\partial \tilde z^2}
+ (1-\lambda)\frac{\partial^2}{\partial \tilde \rho^2}
+ \frac{1}{\tilde \rho}\frac{\partial}{\partial \tilde \rho}
+ \Omega^2
\right]\phi(\tilde \rho,\tilde z)=0.
\label{eq:App_operator}
\end{equation}
We seek separated solutions of the form
\begin{equation}
\phi(\tilde \rho,\tilde z)=e^{i\tilde\beta \tilde z}R(\tilde \rho),
\label{eq:App_sep}
\end{equation}
which gives the radial equation
\begin{equation}
(1-\lambda)R''(\tilde \rho)+\frac{1}{\tilde \rho}R'(\tilde \rho)
+\left(\Omega^2-\tilde \beta^2\right)R(\tilde \rho)=0.
\label{eq:App_radialR}
\end{equation}
Introducing the separation parameter $\eta$ through
\begin{equation}
\tilde \beta^2(\eta)=\Omega^2-(1-\lambda)\eta^2,
\label{eq:App_beta}
\end{equation}
and writing
\begin{equation}
R(\tilde \rho)=\tilde \rho^{\sigma}u(\eta \tilde \rho),
\qquad
\sigma=\frac{\lambda}{2(\lambda-1)},
\label{eq:App_Rsigma}
\end{equation}
Eq.~\eqref{eq:App_radialR} reduces to
\begin{equation}
u''(x)+\frac{1}{x}u'(x)+\left(1-\frac{\nu_0^2}{x^2}\right)u(x)=0,
\qquad
x=\eta \tilde \rho,
\label{eq:App_bessel}
\end{equation}
with
\begin{equation}
\nu_0=\frac{|\lambda|}{2(1-\lambda)}.
\label{eq:App_nu0}
\end{equation}
Equation~\eqref{eq:App_bessel} is the Bessel equation, so the radial sector is spanned by
\begin{equation}
R_\eta(\tilde \rho)=\tilde \rho^\sigma J_{\pm \nu_0}(\eta \tilde \rho).
\label{eq:App_basispm}
\end{equation}

As discussed in the main text, for $\lambda<0$ one has $\sigma=\nu_0>0$, so the small-argument limits
\begin{equation}
J_{\pm \nu_0}(x)\sim x^{\pm \nu_0}
\qquad (x\to 0)
\end{equation}
imply
\begin{equation}
\tilde \rho^\sigma J_{\nu_0}(\eta \tilde \rho)\sim \tilde \rho^{2\nu_0},
\qquad
\tilde \rho^\sigma J_{-\nu_0}(\eta \tilde \rho)\sim \tilde \rho^{0}.
\label{eq:App_axisbranch}
\end{equation}
Therefore the branch compatible with bright on-axis entrance profiles is
\begin{equation}
J_\lambda(x)=
\begin{cases}
J_{\nu_0}(x), & \lambda\ge 0,\\[4pt]
J_{-\nu_0}(x), & \lambda<0.
\end{cases}
\label{eq:App_Jlambda}
\end{equation}

For each fixed $\eta$, Eqs.~\eqref{eq:App_sep} and \eqref{eq:App_basispm} provide a separated mode
\[
\phi_\eta(\tilde\rho,\tilde z)
=
\tilde\rho^\sigma
J_\lambda(\eta\tilde\rho)\,
e^{i\tilde\beta(\eta)\tilde z}.
\]
Since $\eta$ labels a continuous family of such modes, the general forward-propagating axisymmetric solution is obtained by superposing them. The measure $\eta\,d\eta$ is the natural one associated with the Bessel completeness relation used below. Hence the field can be expanded as
\begin{equation}
\phi(\tilde \rho,\tilde z)
=
\tilde \rho^\sigma
\int_0^\infty d\eta\,
\eta\,
A(\eta)\,
e^{i\tilde\beta(\eta)\tilde z}
J_\lambda(\eta \tilde \rho),
\label{eq:App_forward}
\end{equation}
which is the spectral reconstruction used in the numerical analysis. At the launch plane $\tilde z=0$ this becomes
\begin{equation}
\Phi_{\rm in}(\tilde \rho)
=
\tilde \rho^\sigma
\int_0^\infty d\eta\,
\eta\,
A(\eta)\,
J_\lambda(\eta \tilde \rho).
\label{eq:App_launch}
\end{equation}
Defining
\begin{equation}
\Psi(\tilde \rho)=\tilde \rho^{-\sigma}\Phi_{\rm in}(\tilde \rho),
\label{eq:App_Psi}
\end{equation}
one obtains the Hankel-type transform pair
\begin{equation}
\Psi(\tilde \rho)
=
\int_0^\infty d\eta\,\eta\,A(\eta)\,J_\lambda(\eta\tilde \rho),
\label{eq:App_hankel1}
\end{equation}
together with the inverse relation
\begin{equation}
A(\eta)
=
\int_0^\infty d\tilde \rho\,
\tilde \rho\,
J_\lambda(\eta \tilde \rho)\,
\Psi(\tilde \rho),
\label{eq:App_hankel2}
\end{equation}
which follows from the standard Bessel orthogonality relation
\begin{equation}
\int_0^\infty d\eta\,\eta\,
J_\lambda(\eta \tilde \rho)\,
J_\lambda(\eta \tilde \rho')
=
\frac{\delta(\tilde \rho-\tilde \rho')}{\tilde \rho}.
\label{eq:App_orth}
\end{equation}
Since the entrance field is truncated by the finite aperture, $\Phi_{\rm in}(\tilde \rho)=0$ for $\tilde \rho>1$, the integral reduces to
\begin{equation}
A(\eta)
=
\int_0^1 d\tilde \rho\,
\tilde \rho^{\,1-\sigma}
J_\lambda(\eta \tilde \rho)\,
\Phi_{\rm in}(\tilde \rho).
\label{eq:App_Aeta1}
\end{equation}
Using
\begin{equation}
1-\sigma=\sigma+\frac{1}{1-\lambda},
\label{eq:App_sigmaidentity}
\end{equation}
this can be written as
\begin{equation}
A(\eta)
=
\int_0^1 d\tilde \rho\,
\tilde \rho^{\,\sigma+\frac{1}{1-\lambda}}
J_\lambda(\eta \tilde \rho)\,
\Phi_{\rm in}(\tilde \rho),
\label{eq:App_Aeta2}
\end{equation}
which is the form used in the main text.

The connection with the Green kernel can now be read directly. In Eq.~\eqref{eq:App_green_m0},
the factor $J_{\nu_0}(\chi \rho_<) H^{(1)}_{\nu_0}(\chi \rho_>)$ makes explicit
the regular--outgoing radial structure of the anisotropic propagator. When
the problem is formulated instead as forward evolution from a prescribed
launch-plane profile, the same axisymmetric propagation problem is expressed
in terms of a regular radial basis $J_\lambda(\eta \tilde\rho)$, a forward
longitudinal phase factor $e^{i\tilde\beta(\eta)\tilde z}$, and a spectral
amplitude $A(\eta)$ fixed by the entrance data. In this sense, Eqs.~\eqref{eq:App_forward} and \eqref{eq:App_Aeta2} provide the initial-value representation used in the numerical
implementation, while the Green kernel retains the formal propagator
structure of the problem. For this reason, the Green-kernel representation is
kept in the main text as the natural formal description of the propagator,
whereas the spectral form is adopted for the numerical implementation.

\bibliographystyle{apsrev4-2}
\bibliography{PTII_complete_v4_refined}

\end{document}